\documentclass[aps,preprintnumbers,superscriptaddress,nofootinbib,12pt]{revtex4-1}
\usepackage[utf8]{inputenc}
\usepackage{color}
\usepackage{amsfonts,amsmath,amssymb}
\usepackage{graphicx}
\graphicspath{ {./figure/} }
\usepackage{bm}
\usepackage[pdfencoding=auto]{hyperref}
\hypersetup{colorlinks = true,
           allcolors = blue}
\usepackage{bookmark}
\usepackage{color}
\usepackage{epstopdf}
\usepackage{appendix}
\usepackage{ulem}
\allowdisplaybreaks

\newcommand{\itp}{\affiliation{CAS Key Laboratory of Theoretical Physics, Institute of Theoretical Physics, Chinese Academy of Sciences,  Zhong Guan Cun East Street 55, Beijing 100190, China}}

\newcommand{\ucas}{\affiliation{School of Physical Sciences, University of Chinese Academy of Sciences, Beijing 100049, China}}
\renewcommand{\sout}{\bgroup \color[rgb]{1,0,0} \ULdepth=-.5ex \ULset}

\newcommand{\der}{\mathrm{d}}
\newcommand{\be}{\begin{equation}}
\newcommand{\ee}{\end{equation}}
\newcommand{\ba}{\begin{eqnarray}}
\newcommand{\ea}{\end{eqnarray}}

\newcommand{\Lag}{\mathcal{L}}

\newcommand{\Tr}[1]{\left\langle #1 \right\rangle}
\newcommand{\mM}{\mathcal{M}}

\begin{document}

\title{Triangle singularities in $J/\psi\rightarrow\eta\pi^0\phi$  and  $\pi^0\pi^0\phi$}
\author{Hao-Jie Jing}
\email{jinghaojie@itp.ac.cn}
\itp
\ucas
\author{Shuntaro Sakai}
\email{shsakai@itp.ac.cn}
\itp
\author{Feng-Kun Guo}
\email{fkguo@itp.ac.cn}
\itp
\ucas
\author{Bing-Song Zou}
\email{zoubs@itp.ac.cn}
\itp
\ucas
\affiliation{Central South University, Hunan 410083, China}

\date{\today}
\begin{abstract}
The BESIII Collaboration recently reported the observation of the $a_0(980)^0-f_0(980)$ mixing in the isospin breaking decay $J/\psi\to \eta\pi^0\phi$. In the Dalitz plot for that decay with the $\eta$ reconstructed from two photons, there is a band around $1.4$~GeV on the $\pi^0 \phi$ distribution.
In general, this peak can be due to a resonance or a kinematic effect.
In this paper, we study the effects of a set of $K^*K\bar K$ triangle diagrams, and show that due to triangle singularities such diagrams can lead to a peak around 1.4~GeV in the $\pi^0\phi$ invariant mass distribution, which is a model-independent conclusion. The Dalitz plot induced by such a mechanism has a feature consistent with the BESIII observation, namely events along the band accumulate at both ends close to the Dalitz plot boundary.
The effect of the same mechanism on the  $J/\psi\to \pi^0\pi^0\phi$ and $J/\psi \to \eta\pi^0 K^+K^-$ decays are also investigated. We suggest to take more data for the $J/\psi\to \eta\pi^0\phi\to \eta\pi^0 K^+ K^-$ and check whether the structure around $1.4$~GeV persists for the $K^+K^-$ invariant mass away from the $\phi$ mass region. This is crucial for understanding whether the band is due to triangle singularities or due to a resonance. Were it the latter, the band should remain while it would not if it is due to the former.

\end{abstract}

\maketitle
\section{Introduction}
The BESIII Collaboration reported observations of the $a_0(980)^0-f_0(980)$ mixing in the processes $J/\psi \rightarrow \phi f_0(980) \rightarrow \phi a_0(980)^0 \rightarrow \phi\,\pi^0\eta$ and $\chi_{c1} \rightarrow a_0(980)^0\pi^0 \rightarrow f_0(980)\pi^0 \rightarrow \pi^+\pi^-\pi^0$~\cite{Ablikim:2010aa,Ablikim:2018pik}.
The isospin of the $a_0(980)^0$ is 1 and that of the $f_0(980)$ is 0, and thus their mixing breaks isospin. Both of the two resonances couple strongly to the $K\bar K$ meson pair~\cite{Tanabashi:2018oca}, so that the mixing can happen through their coupling to the $K\bar K$ intermediate states and the mass difference between the charged and neutral kaons gives the isospin breaking. Then the mixing probability depends crucially on the coupling strengths of the $a_0(980)$ and $f_0(980)$ to the $K\bar K$. Such a mixing mechanism was first proposed in the late 1970s~\cite{Achasov:1979xc}, and was suggested to contain important information to clarify the nature of these two mesons. That is because the effective couplings of these mesons to the kaons can be related to their probabilities to be $K\bar K$ molecules (see Refs.~\cite{Baru:2003qq,Sekihara:2014kya,Guo:2015daa,Kamiya:2016oao} for discussions based on extensions of Weinberg's compositeness relations~\cite{Weinberg:1965zz}).
The $J/\psi\rightarrow\eta\pi^0\phi$ reaction as a probe of the $a_0(980)^0-f_0(980)$ mixing was suggested in Ref.~\cite{Wu:2007jh}, and was further analyzed using unitarized chiral approaches in Refs.~\cite{Hanhart:2007bd,Roca:2012cv}.

In the Dalitz plot of the $J/\psi \rightarrow\eta\pi^0\phi$ process measured by the BESIII Collaboration (see Fig.~1 of Ref.~\cite{Ablikim:2018pik}), it is clear that there is a peak near the $K \bar{K}$ threshold on the $\eta\pi^0$ distribution which can be interpreted as the $a_0(980)^0-f_0(980)$ mixing.
In addition, there is also a clear accumulation of events for the $\pi^0 \phi$ invariant mass being around $1.4$~GeV, which would be a peak if the Dalitz plot is projected to the $\pi^0 \phi$ invariant mass distribution.
In general, this peak can be due to a resonance with a mass about 1.4~GeV (isospin $I=1$ or $I=0$) or a kinematic effect. From the first point of view, because the $S$-wave $\pi^0 \phi$ has $J^{PC} = 1^{+-}$ and this decay breaks isospin symmetry, that peak may be related to the isovector resonance claimed in Ref.~\cite{Bityukov:1986yd} or the isoscalar $h_1(1380)$ whose mass is about $1.41$~GeV~\cite{Ablikim:2015lnn,Ablikim:2018ctf,Tanabashi:2018oca}.\footnote{Some discussions on the spectrum of axial-vector mesons with $q\bar{q}$ constituents can be found in, e.g., Refs.~\cite{Godfrey:1985xj,Cheng:2011pb}, and a possible description of the $h_1(1380)$ as a $K^*\bar{K}$ molecule was  proposed in Ref.~\cite{Roca:2005nm}.}
From the second point of view, because this is an isospin breaking process and that peak position about $1.4$~GeV is near the threshold of $K^*\bar{K}$, similar to the enhancement of the isospin violation by triangle singularity as seen in the $\eta(1405/1475) \rightarrow \pi^0 f_0(980)$~\cite{Wu:2011yx,Aceti:2012dj,Wu:2012pg,Achasov:2015uua,Du:2019idk}, triangle diagrams involving $K^*\bar{K}K$ intermediate states can also lead to a peak and a significant isospin breaking enhancement here (see also Sec.~VI.A.4 of Ref.~\cite{Guo:2017jvc} and references therein for the discussion of the possible role of triangle singularity on the $h_1(1380)$).

\begin{figure}[tb]
    \centering
    \includegraphics[width=0.88\textwidth]{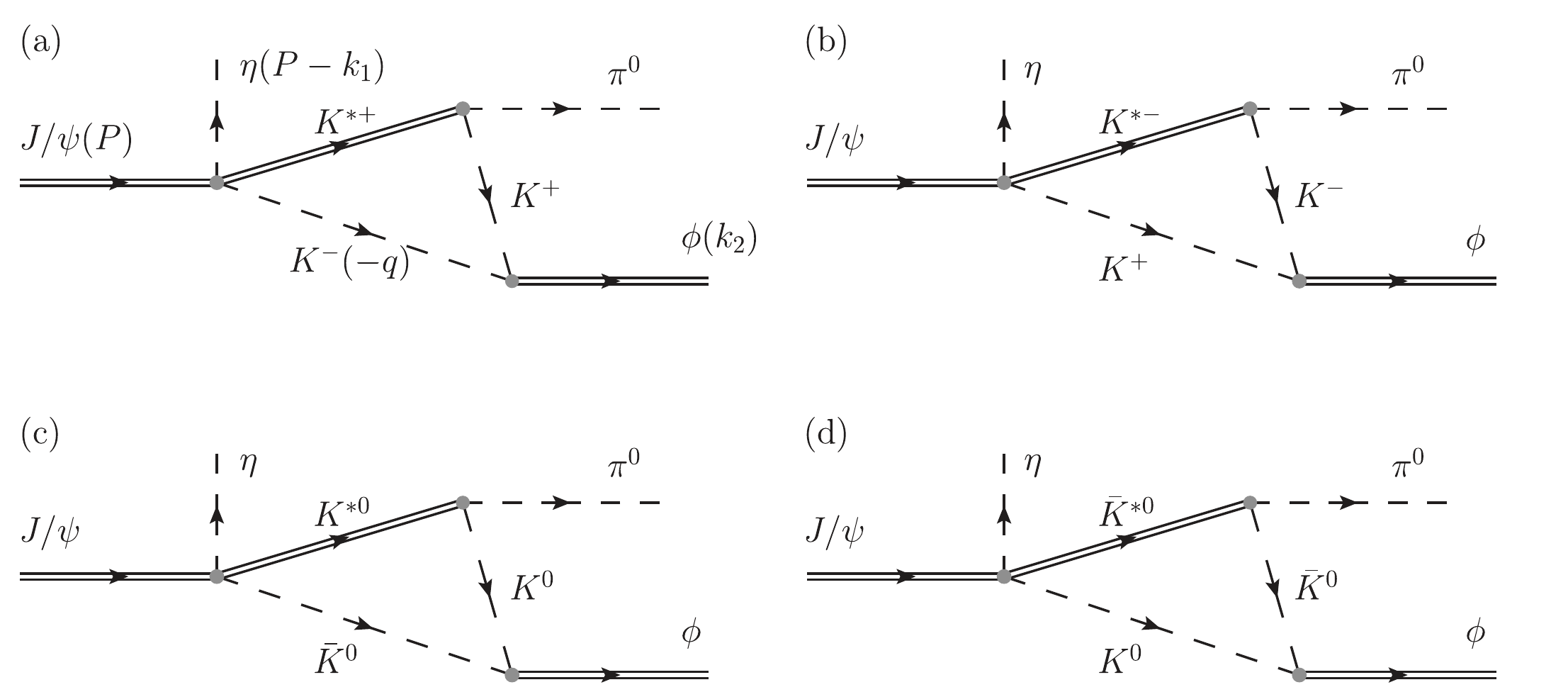}
    \caption{Triangle diagrams in the $J/\psi$ decay into $\eta\pi^0\phi$ that can produce a peak at $m_{\pi^0\phi}\simeq1.4$~GeV without a resonance.}
    \label{FIG1to3}
\end{figure}

Triangle singularity is the leading Landau singularity~\cite{Landau:1959fi} of a triangle diagram, and it depends crucially on the kinematics. To be more specific, when all of the three intermediate particles in a triangle diagram are on shell, moving collinearly, and all of the interaction vertices satisfy the energy-momentum conservation~\cite{Coleman:1965xm}, the physical amplitude has a logarithmic triangle singularity leading to a peak in the invariant mass distributions (see also Ref.~\cite{Bayar:2016ftu} for an intuitive picture of triangle singularity with a reformulation based on a diagrammatic approach).
The production of $\pi^0\phi$ in the reaction $\pi^- p \to \pi^0\phi n$ from the $K^*\bar{K}K$ triangle singularity mechanism has been studied in Ref.~\cite{Achasov:1989ma}.
In the $J/\psi\rightarrow\eta\pi^0\phi$ reaction, we focus on the diagrams that can lead to a peak around the $K^*\bar K$ threshold, about $1.4$~GeV, in the $\pi^0\phi$ distribution, and thus consider the triangle diagrams shown in Fig.~\ref{FIG1to3}. The triangle singularity for each of these diagrams is at the physical boundary when the kinematics is such that the following processes happen:
first, the $J/\psi$ decays into $\eta K^*\bar{K}$ (the charge-conjugated $\bar K^* K$ diagrams are also included, see Fig.~\ref{FIG1to3}), and the $K^*$ decays into $\pi K$ subsequently;
then the $K$ moves in the same direction with the $\bar K$ and catches up with it, and the $K\bar{K}$ pair finally forms the $\phi$.
Triangle singularity effects of the $K^* K\bar K$ loops have been studied for processes with the $K\bar K$ pair forming an $f_0(980)$ or $a_0(980)$ instead of a $\phi$ in Refs.~\cite{Wu:2011yx,Wu:2012pg,Aceti:2012dj,Ketzer:2015tqa,Achasov:2015uua,Aceti:2016yeb,Debastiani:2016xgg,Sakai:2017iqs,Liang:2017ijf,Liang:2019jtr,Du:2019idk}.

This paper is organized as follows. In Section~\ref{sec:form}, we set up the formalism for calculating the decays $J/\psi \rightarrow \eta\pi^0\phi$,  $J/\psi \rightarrow \pi^0\pi^0\phi$, and $J/\psi \rightarrow \eta\pi^0 K\bar K$. For the last process, the $K\bar K$ can be either isovector or isoscalar and the isoscalar $K\bar K$ final state interaction (FSI) is taken into account using the inverse amplitude method. As will be shown in Section~\ref{sec:results}, by considering the triangle diagrams shown in Fig.~\ref{FIG1to3} for the $J/\psi \rightarrow \eta\pi^0\phi$, a peak appears in the $\pi^0\phi$ invariant mass distribution around $1.4$~GeV by virtue of triangle singularity, which might be the physics behind the band at $m_{\pi^0\phi}\simeq1.4$~GeV in the Dalitz plot measured by the BESIII Collaboration as they have similar gross features.
The results for the $\pi^0\phi$ distribution of the $J/\psi\rightarrow\pi^0\pi^0\phi$ reaction will also be shown, and the peak induced by the triangle diagrams is much broader since this is an isospin symmetry conserving reaction.
A brief summary is given in Section~\ref{sec:summary}. Some decay amplitudes and  a derivation of the generic $n$-body phase space are relegated in Appendix~\ref{app:mM} and Appendix~\ref{app:PSV}, respectively.

\section{Formalism}
\label{sec:form}

\subsection{\texorpdfstring{\bm{$J/\psi \rightarrow \eta \pi^0 \phi $}}{Jpepiphi}}

The diagrams we consider for the $J/\psi \to \eta\pi^0\phi$ are shown in Fig.~\ref{FIG1to3}. To obtain the decay amplitude, we use the following effective Lagrangian for the  $J/\psi\rightarrow\eta K^*\bar{K}$, $K^*\rightarrow\pi K$, and $K\bar{K}\rightarrow\phi$ vertices:
\begin{align}
  \Lag_\text{int} =&\, \Lag_{JVPP}+\Lag_{VPP},  \nonumber\\
  \Lag_{JVPP} =&\, g_1J_\mu\Tr{V^\mu P P},  \label{Lagrangian}\\
  \Lag_{VPP} =&-i g_2 \Tr{V^\mu[P,\partial_\mu P]}, \nonumber
\end{align}
where $g_1$ and $g_2$ are coupling constants, and $J_\mu$ is the field operator of the $J/\psi$, which as a charmonium is a light-flavor SU(3) singlet state. The light pseudoscalar meson octet and vector meson nonet matrices are denoted by $P$ and $V_\mu$, respectively,
\ba
  P = \frac{1}{\sqrt{2}}\begin{pmatrix} \pi^0+\frac{1}{\sqrt{3}}\eta&\sqrt2 \pi^+&\sqrt2 K^+ \\
                    \sqrt2 \pi^-&-\pi^0+\frac{1}{\sqrt{3}}\eta&\sqrt2 K^0\\
                    \sqrt2 K^-&\sqrt2 \bar{K}^0&-\frac{2}{\sqrt{3}}\eta\end{pmatrix},\quad
  V_\mu = \frac{1}{\sqrt{2}}\begin{pmatrix}\rho^0+\omega&\sqrt2 \rho^+&\sqrt2 K^{*+} \\
                    \sqrt2 \rho^-&-\rho^0+\omega&\sqrt2 K^{*0}\\
                    \sqrt2 K^{*-}&\sqrt2 \bar{K}^{*0}&\sqrt{2}\phi\end{pmatrix}_\mu ,~~
\ea
where the ideal mixing between the $\omega$ and $\phi$ is assumed such that the valence quarks for them are $(u\bar u+d\bar d)/\sqrt{2}$ and $s\bar s$, respectively.
The $VPP$ Lagrangian\footnote{one can check that this $VPP$ Lagrangian is phenomenologically good at explaining the strong interaction decay width of light vector mesons.} in Eq.~\eqref{Lagrangian} may be obtained in the hidden-local symmetry framework (see, e.g., Refs.~\cite{Bando:1984ej,Meissner:1987ge,Klingl:1996by} and references therein),
and the effective Lagrangian for the $J/\psi VPP$ vertex has been used in Refs.~\cite{Meissner:2000bc,Roca:2004uc,Liang:2019vhf} to study the $J/\psi$ decays into one vector and two pseudoscalar mesons.
A term proportional to $J_\mu\Tr{V^\mu }\Tr{PP}$ involves two flavor traces; thus, it is suppressed by a factor of $1/N_c$, with $N_c$ being the number of colors, compared to Eq.~\eqref{Lagrangian} and will not be considered in this calculation. Indeed, this term was considered in Ref.~\cite{Liang:2019vhf} and was found relatively small.

The decay $J/\psi\rightarrow\eta\pi^0\phi$ breaks isospin symmetry. In the mechanism considered in this study, the isospin breaking comes from the mass differences between the charged and neutral $K$ and $K^*$ mesons in the triangle loops shown in Fig.~\ref{FIG1to3}.
From the Lagrangian given above, one can obtain the amplitude for each vertex:
\begin{alignat}{5}
&-it_{J/\psi,K^{*+}K^-\eta} = i \frac{1}{\sqrt{6}} g_1        \epsilon^\mu_{J/\psi}\epsilon^{*\nu}_{K^{*+}}g_{\mu\nu},&
&-it_{J/\psi,K^{*0}\bar{K}^0\eta} = i \frac{1}{\sqrt{6}} g_1                    \epsilon^\mu_{J/\psi}\epsilon^{*\nu}_{K^{*0}}g_{\mu\nu}, \nonumber\\
&-it_{J/\psi,K^{*-}K^+\eta} = i \frac{1}{\sqrt{6}} g_1                    \epsilon^\mu_{J/\psi}\epsilon^{*\nu}_{K^{*-}}g_{\mu\nu},&
&-it_{J/\psi,\bar{K}^{*0}K^0\eta} = i \frac{1}{\sqrt{6}} g_1              \epsilon^\mu_{J/\psi}\epsilon^{*\nu}_{\bar{K}^{*0}}g_{\mu\nu},  \nonumber\\
&-it_{K^{*+},K^+\pi^0} = -i \frac{\sqrt{2}}{2} g_2                     \epsilon^\mu_{K^{*+}}(p_{K^+}-p_{\pi^0})_\mu,&
~~&-it_{K^{*-},K^-\pi^0} = -i \frac{\sqrt{2}}{2}g_2 \epsilon^\mu_{K^{*-}}(p_{\pi^0}-p_{K^-})_\mu, \label{eq:ampjpkkst}\\
&-it_{K^{*0},K^0\pi^0} = -i \frac{\sqrt{2}}{2}g_2 \epsilon^\mu_{K^{*0}}(p_{\pi^0}-p_{K^0})_\mu,&
&-it_{\bar{K}^{*0},\bar{K}^0\pi^0} = -i \frac{\sqrt{2}}{2}g_2                   \epsilon^\mu_{\bar{K}^{*0}}(p_{\bar{K}^0}-p_{\pi^0})_\mu, \nonumber\\
&-it_{\phi,K^+K^-} = -ig_2 \epsilon^\mu_\phi(p_{K^+}-p_{K^-})_\mu,&
&-it_{\phi,K^0\bar{K}^0} = -ig_2 \epsilon^\mu_\phi(p_{K^0}-p_{\bar{K}^0})_\mu.\nonumber
\end{alignat}
The coupling constant $g_2$ for the coupling of $\phi\to K\bar{K}$, $K^*\to\pi K$, and $\bar{K}^*\to \pi\bar{K}$ can be fixed from reproducing the observed $\phi$ and $K^*$ widths~\cite{Tanabashi:2018oca}: $g_2\simeq4.5$,
The parameter $g_1$ is unknown. However, since we only care about the shape and relative size of the invariant mass distributions, we may set $g_1=1$.\footnote{In fact, the value of $g_2$ is also irrelevant for the characteristic feature of the nontrivial structures in question.}

With Eq.~\eqref{eq:ampjpkkst}, the amplitude for a given diagram in Fig.~\ref{FIG1to3} can be written as follows:
\begin{align}
    -i{\cal M}_{id} &=-{\frac{g_1g_2^2}{2\sqrt{3}}}\epsilon_{J/\psi}^\mu \epsilon_{\phi}^\nu
             \int \frac{\mathrm{d}^4 q}{(2\pi)^4} \frac{\left[-g_{\mu \lambda} + {(q+k_1)_\mu(q+k_1)_\lambda}/{m_{K_{id}^*}^2}\right](q+2k_2-k_1)^\lambda (2q+k_2)_\nu}
            {(q^2-m_{K_{id}}^2+i\epsilon)[(q+k_1)^2-m_{K_{id}^*}^2+i\epsilon][(q+k_2)^2-m_{K_{id}}^2+i\epsilon]} \nonumber\\
            &\equiv -{\frac{g_1g_2^2}{2\sqrt{3}}} \epsilon_{J/\psi}^\mu \epsilon_{\phi}^\nu {\cal M}^{id}_{\mu \nu}\,,
\label{AmpOf1to3}
\end{align}
where the index $id=C (N)$ corresponding to the process with charged (neutral) intermediate particles.
The details of $\mM_{id}^{\mu\nu}$ are given in Appendix~\ref{app:mM}.
Adding these charged and neutral loop amplitudes with appropriate phases,
the total amplitude is given by
\begin{align}
  \mM_{J/\psi\rightarrow\eta\pi^0\phi}=2(\mM_C-\mM_N).
  \label{eq:mtot}
\end{align}
The factor $2$ is to take account of the charge conjugated contributions, i.e.,
the amplitudes for Fig.~\ref{FIG1to3}\,(a,\,c)  give the same contribution as those for Fig.~\ref{FIG1to3}\,(b,\,d).
As one can see from Eq.~\eqref{AmpOf1to3}, the amplitudes $\mM_C$ and $\mM_N$ differ only by the masses of the intermediate mesons $K$ and $K^*$,
and in the isospin limit, which is realized by using identical masses for the mesons in the same isospin multiplet, $\mM_C$ and $\mM_N$ in  Eq.~\eqref{eq:mtot} would exactly cancel each other.
We calculate this amplitude in the center-of-mass (CM) frame of the $\pi^0$ and $\phi$ pair, and choose the $\pi^0$ momentum direction as the $z$-direction.

The differential width of this process can be written as (see Appendix~\ref{app:PSV} for details of the phase space factor)
\be
\mathrm{d}\Gamma = \frac{1}{(2\pi)^52^4m_{J/\psi}^2}\frac{1}{3}\sum_\text{spin}|\mM|^2|{\mathbf p}_1||{\mathbf p}^{*}_2|\mathrm{d}\Omega_1\mathrm{d}\Omega_2^*\mathrm{d}m_{23},
\label{dGammaOf1to3}
\ee
where the quantities marked with $*$ are evaluated in the CM frame of particle 2 and particle 3 in the final states, and the momentum $(|\mathbf{p}_1|, \Omega_1)$ is the momentum of particle 1 in the rest frame of the decay particle.
Since we are interested in the invariant mass distribution of $\pi^0\phi$, we choose $\eta$ as particle 1, $\pi^0$ as  particle 2, and $\phi$ as particle 3.
Then the differential width is given by
\be
    \frac{\der\Gamma_{J/\psi\rightarrow \eta\pi^0\phi}}{\der m_{\pi^0\phi}}= \frac{|{\mathbf p}_\eta||{\mathbf p}^{*}_{\pi^0}|}{(2\pi)^52^4m_{J/\psi}^2}\frac{1}{3} \int \der \Omega_\eta
    \der\Omega_{\pi^0}^{*}\sum_\text{spin}|\mM_{J/\psi\rightarrow \eta\pi^0\phi}|^2,\label{eq:dgamdm}
\ee
with  $\sum_\text{spin}$ summing over polarizations of $J/\psi$ and $\phi$ and the amplitude $\mM_{J/\psi\rightarrow \eta\pi^0\phi}$ given in Eq.~\eqref{eq:mtot}.

\subsection{\texorpdfstring{\bm{$ J/\psi \rightarrow \eta \pi^0 K \bar{K} $}}{JpepKKbar}}

\begin{figure}
    \centering
    \includegraphics[width=0.88\textwidth]{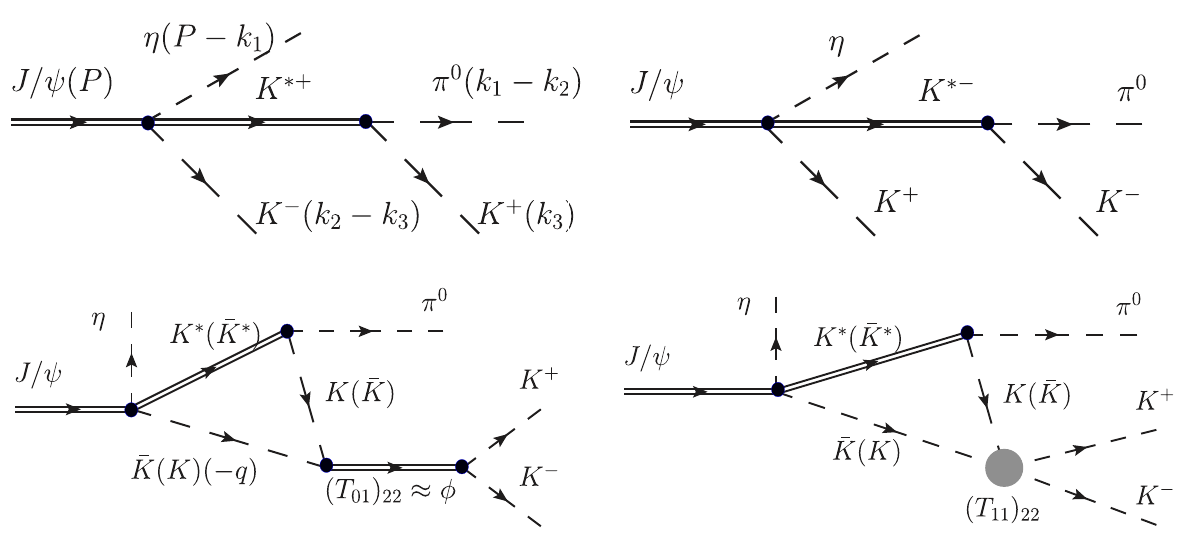}
    \caption{Feynman diagrams for the $J/\psi$ decay into $\eta\pi^0 K^+ K^-$ considered in this work. In the triangle diagrams, both the charged and neutral $K$ and $K^*$ mesons are taken into account as in Fig.~\ref{FIG1to3}. In the second line, $(T_{01})_{22}$ and $(T_{11})_{22}$ refer to the $P$-wave $K\bar K\to K\bar K$ scattering amplitudes with isospin $I=0$ and $I=1$, respectively. }
    \label{FIG1to4}
\end{figure}

In the experimental measurement, the $\phi$ is reconstructed from its decays into the $K^+ K^-$ final states with the $K^+ K^-$ invariant mass in the region $[m_\phi-10~\text{MeV}, m_\phi+10~\text{MeV}]$. For a direct comparison with the experimental data, we need to take into account the subsequential decay of the $\phi$ into $K^+K^-$ which means the $\eta \pi^0 K^+ K^-$ in the final state.
In this case, since the $K^*$ decay into $K\pi$ with an almost 100\% branching fraction, tree diagrams should be included in addition to the triangle loop ones.
The diagrams for this process considered here are shown in Fig.~\ref{FIG1to4}.
The triangle diagrams involve the $K\bar{K}\rightarrow K\bar{K}$ final state interaction.
We consider the $P$-wave $K\bar{K}$ scattering amplitude with quantum numbers $(I,J)=(0,1)$ and $(1,1)$.
For simplicity, only the contribution of the $\phi$ meson is considered in the $(0,1)$ channel.
For the $K\bar{K}$ $(I,J)=(1,1)$ amplitude, we  employ the inverse amplitude method as developed in Ref.~\cite{PhysRevD.59.074001}, which is a unitary extension of chiral perturbation theory and can describe the meson-meson scattering data up to 1.2~GeV.
In this approach, the unitarized $T$-matrix for a given partial wave can be written as
\be
\centering
T = T_2(T_2-T_4-T_2GT_2)^{-1}T_2,
\ee
where $T_2$ is the partial wave scattering amplitude from the leading order chiral Lagrangian, $T_4$ is the polynomial tree-level amplitude from the next-to-leading order chiral Lagrangian, and $G$ is a diagonal matrix given by the loop integrals with two meson propagators.
One can find the specific form of the $T$-matrix in Ref.~\cite{PhysRevD.59.074001}.
Here, we use the isospin symmetric $K\bar{K}\rightarrow K\bar{K}$ amplitude to concentrate on the effect of triangle singularities.
The amplitudes for the  $K\bar{K}\to K\bar K$ FSI with quantum numbers $(I,J)=(0,1)$ and $(1,1)$ are written as:
\begin{align}
-i t_{(0,1)} &= -i g^2_2 \frac{(-g^{\mu\nu}+p^\mu_\phi p^\nu_\phi/m^2_\phi)}{m^2_{K \bar{K}}-m^2_{\phi}+i m_\phi \Gamma_\phi} (p^\text{in}_K - p^\text{in}_{\bar K})_\mu (p^\text{out}_K - p^\text{out}_{\bar K})_\nu \notag\\
&= i\frac{g_2^2}{m^2_{K \bar{K}}-m^2_{\phi} + i m_\phi \Gamma_\phi}(p^\text{in}_K-p^\text{in}_{\bar K})^\mu  (p^\text{out}_K-p^\text{out}_{\bar K})_\mu \notag\\
&\equiv -i\tilde{t}_{(0,1)}(p^\text{in}_K-p^\text{in}_{\bar K})^\mu  (p^\text{out}_K-p^\text{out}_{\bar K})_\mu, \\
-i t_{(1,1)} &= -i 3(T_{11})_{22} \cos \theta  \notag\\
&= -i 3(T_{11})_{22}  \frac{(p^\text{in}_K-p^\text{in}_{\bar K})^\mu  (p^\text{out}_K-p^\text{out}_{\bar K})_\mu}{4 q^\text{in}q^\text{out}}\label{eq:t11-med}  \notag\\
&\equiv -i\tilde{t}_{(1,1)}(p^\text{in}_K-p^\text{in}_{\bar K})^\mu  (p^\text{out}_K-p^\text{out}_{\bar K})_\mu,
\end{align}
where $m_\phi=1019.5$~MeV and $\Gamma_\phi=4.2$~MeV are the mass and width of the $\phi$ meson, respectively~\cite{Tanabashi:2018oca}, $p_\phi=p^\text{out}_K+p^\text{out}_{\bar K}$, and $\theta$ is the angle between the 3-momenta of $K$ mesons in the loop and final state in the $K \bar{K}$ CM frame.
The momenta $p^\text{in}_K$ and $p^\text{out}_K$ ($p^\text{in}_{\bar{K}}$ and $p^\text{out}_{\bar{K}}$) denote the $K$ $(\bar{K})$ momenta in the loop and final state, respectively, and $q^\text{in}, q^\text{out}$ are the magnitudes of the corresponding CM momenta.
The $T$-matrix $T_{11}$ is a two-channel matrix with the quantum numbers $(I=1,J=1)$, whose elements $(T_{11})_{11}$, $(T_{11})_{12}$ and $(T_{11})_{22}$ correspond to the $P$-wave scattering amplitudes for $\pi\pi\rightarrow\pi\pi$, $\pi\pi\rightarrow K\bar{K}$ and $K\bar{K}\rightarrow K\bar{K}$, respectively.
The purpose of our calculation is to study the influence of triangle singularity on the $\pi^0K\bar K$ invariant mass distribution, so we focus on the kinematic region where triangle singularity occurs, i.e. for all the intermediate particles being on-shell.
We approximate the momentum $q^\text{in}$ in the denominator of Eq.~\eqref{eq:t11-med} for the $K\bar{K}$ scattering vertex with the  on-shell momentum $q_\text{cm} = q^\text{out}=\sqrt{m_{K\bar K}^2/4-m_K^2}$.\footnote{Because the triangle singularity appears when the internal particles are on-shell, the off-shell contribution that is neglected here just gives non-singular contributions that are smooth in invariant mass distributions.
Then, the structure around $1.4$~GeV to be discussed later would not be changed by this approximation.
Particularly, in the isospin-violating $J/\psi\to\eta\pi^0\phi$ reaction, the process is driven by the triangle singularities,
and the non-singular contribution from the off-shell part would have large cancellation between the charged and neutral meson loops due the isospin symmetry.}
Before we apply this formula, we need to transform the particle basis from the isospin space to the charged space:
\be
\begin{split}
   &|K\bar{K}\rangle_{I=0} = -\frac{1}{\sqrt{2}}|K^+K^-\rangle-\frac{1}{\sqrt{2}}|K^0\bar{K}^0\rangle \,, \\
   &|K\bar{K}\rangle_{I=1,I_3=0} = -\frac{1}{\sqrt{2}}|K^+K^-\rangle+\frac{1}{\sqrt{2}}|K^0\bar{K}^0\rangle\,,
\end{split}
\ee
where the phase convention  $\left|K^{-}\right>=-\left|I=1/2,I_z=-1/2\right>$, the same as that in Ref.~\cite{PhysRevD.59.074001}, is used.
Then, the $J/\psi\rightarrow\eta\pi^0K^+K^-$ amplitudes of the tree and triangle loop diagrams shown in Fig.~\ref{FIG1to4} can be written as follows:
\begin{align}
\begin{split}
-i{\cal M}_\text{tree} &= -i \frac{{1}}{2\sqrt{3}} g_1 g_2 \epsilon_{J/\psi}^\mu\left[\frac{-(p_{\pi^0}-p_{K})_\mu +(p_{\pi^0}+p_{K})_\mu{{(m^2_{\pi^0}-m^2_K)}/{m^2_{K^*}}}}{m^2_{\pi^0 K \bar{K}}-2m_{\pi^0 K\bar{K}}E_{\bar{K}}+m^2_{K}-m^2_{K^*}+i\epsilon}\right.\\
&\left.\hspace{3.5cm}+\frac{-(p_{\pi^0}-p_{\bar{K}})_\mu +(p_{\pi^0}+p_{\bar{K}})_\mu{{(m^2_{\pi^0}-m^2_K)}/{m^2_{K^*}}}}{m^2_{\pi^0 K \bar{K}}-2m_{\pi^0 K \bar{K}}E_{K}+m^2_{K}-m^2_{K^*}+i\epsilon}\right], \\
-i{\cal M}^{I=0}_\text{loop} &=-{\frac{{1}}{2\sqrt{3}}} g_1 g_2\epsilon_{J/\psi}^\mu ({\cal M}^C_{\mu\nu}-{\cal M}^N_{\mu\nu})\tilde{t}_{(0,1)}(p_{\bar{K}}-p_{K})^\nu, \\
    -i{\cal M}^{I=1}_\text{loop} &= -\frac{{1}}{2\sqrt{3}}g_1 g_2\epsilon_{J/\psi}^\mu({{\cal M}^C_{\mu \nu}}+{{\cal M}^N_{\mu \nu}}){\tilde{t}_{(1,1)}(p_{\bar{K}}-p_{K})^\nu},\\
    {\cal M}_\text{tot} &= {\cal M}_\text{tree} + {\cal M}^{I=0}_\text{loop} + {\cal M}^{I=1}_\text{loop}\,.
\end{split}
\label{AmpOf1to4Eta}
\end{align}

Finally, as shown in {Appendix}~\ref{app:PSV},
the differential width of this 4-body decay process can be written as
\be
   \mathrm{d}\Gamma = \frac1{(2\pi)^{8}2^5 m_{J/\psi}^2}\frac{1}{3}\sum_\text{spin} |{\cal M}_\text{tot}|^2 |{\mathbf p}_{1}||{\mathbf p}'_{2}||{\mathbf p}''_{3}|
                   \mathrm{d}\Omega_{1}\mathrm{d}\Omega'_{2}\mathrm{d}\Omega''_{3}\mathrm{d}m_{234}\mathrm{d}m_{34},
\ee
where the momenta $(|{\mathbf p}_{1}|,\Omega_1)$, $(|{\mathbf p}'_{2}|,\Omega'_2)$ and $(|{\mathbf p}''_{3}|,\Omega''_3)$ are evaluated in the rest frame of the decaying particle, the particles $2+3+4$, and the particles $3+4$, respectively. We choose $\eta$ as particle 1,  $\pi^0$ as particle 2,  $K$ as particle 3 and $\bar{K}$ as particle 4.
Then the double differential distribution is given by
\begin{align}
    \frac{\der^2\Gamma_{J/\psi \rightarrow \eta\pi^0K\bar{K}}}{\der m_{\pi^0K\bar{K}}\der m_{K\bar{K}}}= \frac{|{\mathbf p}_{\eta}||{\mathbf p}'_{\pi^0}||{\mathbf p}''_{K}|}{(2\pi)^{8}2^5 m_{J/\psi}^2}\frac{1}{3} \int\mathrm{d}\Omega_{\eta}\mathrm{d}\Omega'_{\pi^0}\mathrm{d}\Omega''_{K}\sum_\text{spin} |{\cal M}_\text{tot}|^2.
    \label{dGammaOf1to4Eta}
\end{align}

\subsection{\texorpdfstring{\bm{$J/\psi\rightarrow \pi^0\pi^0\phi$}}{Jppipiphi}}

Furthermore, we can consider the $J/\psi\rightarrow\pi^0 \pi^0\phi$ process with little additional effort because we only need to replace the $\eta$ by the $\pi^0$ in Fig.~\ref{FIG1to3}.
The difference is that the $J/\psi\rightarrow\eta \pi^0\phi$ breaks isospin symmetry while this process is isospin conserving.
The $J/\psi\rightarrow\pi\pi\phi$ reaction has been studied in Refs.~\cite{Meissner:2000bc,Roca:2004uc,Lahde:2006wr,Liu:2009ub} using the $S$-wave $\pi\pi$ final state interactions to explain the structures in the $\pi\pi$ invariant mass distributions of the $J/\psi\rightarrow\phi\pi\pi$ and $J/\psi \to \omega\pi\pi$~\cite{Augustin:1988ja,Ablikim:2004qna,Ablikim:2004wn}.

From the Lagrangian in Eq.~\eqref{Lagrangian}, we can obtain the following $J/\psi\rightarrow\pi^0K^*\bar{K}$ amplitudes:
\be
\begin{split}
&-it_{J/\psi,K^{*+}K^-\pi^0} = i \frac{\sqrt{2}}{2} g_1         \epsilon^\mu_{J/\psi}\epsilon^{*\nu}_{K^{*+}}g_{\mu\nu}\,, \hspace{1.5cm}
-it_{J/\psi,K^{*0}\bar{K}^0\pi^0} =-i \frac{\sqrt{2}}{2} g_1                    \epsilon^\mu_{J/\psi}\epsilon^{*\nu}_{K^{*0}}g_{\mu\nu}\,, \\
&-it_{J/\psi,K^{*-}K^+\pi^0} = i \frac{\sqrt{2}}{2} g_1                    \epsilon^\mu_{J/\psi}\epsilon^{*\nu}_{K^{*-}}g_{\mu\nu}\,, \hspace{1.5cm}
-it_{J/\psi,\bar{K}^{*0}K^0\pi^0} =- i \frac{\sqrt{2}}{2} g_1              \epsilon^\mu_{J/\psi}\epsilon^{*\nu}_{\bar{K}^{*0}}g_{\mu\nu}\,.  \\
\end{split}
\ee
Due to the difference of the signs from those in Eq.~\eqref{eq:ampjpkkst} for the $J/\psi\rightarrow\eta K^*\bar{K}$ case,
the total amplitude of the $J/\psi\rightarrow\pi^0\pi^0\phi$ is given by
\begin{align}
  \mM_{J/\psi\rightarrow\pi^0\pi^0\phi}=2(\mM_C+\mM_N).
  \label{eq:mtot-ppphi}
\end{align}
One sees that the charged and neutral intermediate states give the same contributions in the isospin limit as expected since the transition preserves isospin symmetry.

The $\pi^0\phi$ mass distribution is obtained from Eq.~\eqref{eq:dgamdm} by replacing $\eta$ with $\pi^0$  and multiplying an additional factor $1/2!$ due to the identical two neutral pions in the $J/\psi\rightarrow\pi^0\pi^0\phi$.

\section{Results}
\label{sec:results}

\subsection{General discussion}

\begin{figure}[t]
    \centering
    \includegraphics[width=0.5\textwidth]{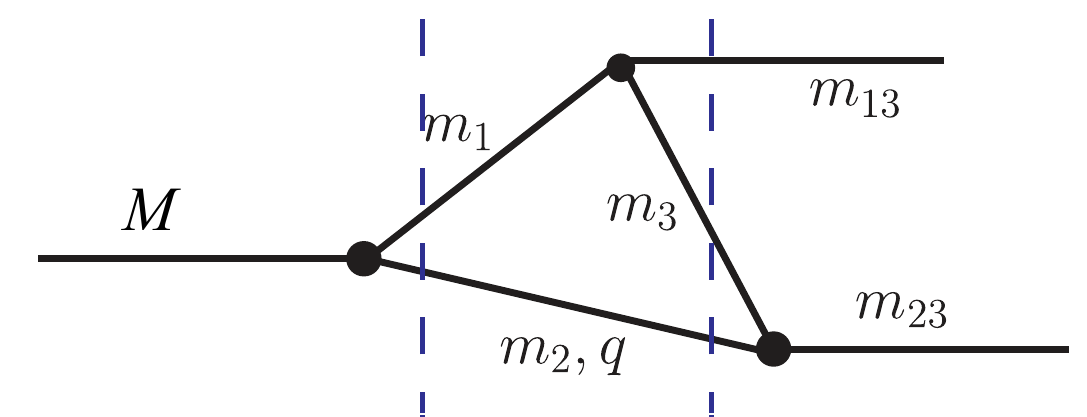}
    \caption{Triangle diagram with the intermediate particles with masses $m_{1,2,3}$. $M,m_{13}$ and $m_{23}$ are the invariant masses of external particles. The two vertical dashed lines denote two cuts.}
    \label{fig:triangle}
\end{figure}

Triangle singularity at the physical boundary can be easily obtained by solving the equation derived in Ref.~\cite{Bayar:2016ftu} which reads (see Fig.~\ref{fig:triangle} for the particle assignment)
\begin{eqnarray}
  q^{}_{\rm on+} = q^{}_{a-}, \quad \text{with}\quad
   q^{}_{{\rm on}+} = \frac1{2 M} \sqrt{\lambda(M^2,m_1^2,m_2^2)},~~ q^{}_{a-} = \gamma \left( \beta \, E_2^* - p_2^* \right),
   \label{eq:ts}
\end{eqnarray}
where $\lambda(x,y,z)=x^2+y^2+z^2-2xy-2yz-2zx$ is the K\"all\'en triangle function, $E_2^*= (m_{23}^2+m_2^2-m_3^2)/(2m_{23})$ and $p_2^*=\sqrt{\lambda(m_{23}^2,m_2^2,m_3^2)}/(2m_{23})$ are the energy and the magnitude of the 3-momentum of particle $m_2$ in the CM frame of $(m_2,m_3)$, $\beta$ is the magnitude of the velocity of $(m_2,m_3)$ system in the CM frame of $(m_1,m_2)$, and $\gamma= 1/{\sqrt{1-\beta^2}}$ is the Lorentz boost factor from the $(m_2,m_3)$ CM frame to the $(m_1,m_2)$ CM frame.

For the triangle singularity to be on the physical boundary, the physical conditions of which are given by the Coleman-Norton theorem~\cite{Coleman:1965xm}, the above equation needs to have a real solution, and all the arguments of the involved square root functions need to be positive. Physically, this means that all the intermediate particles go on shell, moving collinearly, and the $m_3$ particle  from the decay of $m_1$ should be fast enough to catch up with the $m_2$ from the decay of $M$, and then particles $m_2$ and $m_3$ interact like a classical process producing external particle(s) with an invariant mass $m_{23}$. For more discussions, we refer to, e.g., Refs.~\cite{Guo:2015umn,Guo:2017wzr}.

If we fix the the masses of the intermediate states, and the invariant mass $m_{13}$ (corresponding to the $\pi^0$ mass for the question under study), then to have a physical region singularity requires the lower bound for the invariant mass $m_{23}$ to be the $m_2+m_3$ threshold,  and the upper bound can be obtained by solving Eq.~\eqref{eq:ts} with $M=m_1+m_2$, i.e., at the boundary that $m_1$ and $m_2$ can be on shell. As a result, the region of $m_{23}$ for having a triangle singularity in the invariant mass distribution of $M$ at the physical boundary is
\begin{equation}
    m_{23} \in \left[m_2+m_3, \sqrt{ \left[\left(m_1+m_2\right)\left(m_3^2+m_1 m_2\right)-m_2 m_{13}^2\right]/m_1}\,\right].
    \label{eq:range}
\end{equation}

\subsection{\texorpdfstring{\bm{$J/\psi \rightarrow \eta \pi^0 \phi $}}{Jpepiphi}}

We can compute where the logarithmic singularities of the charged and neutral $K^* K\bar K$ triangle loops in Figs.~\ref{FIG1to3} and \ref{FIG1to4} are located using Eq.~\eqref{eq:ts}.
For the $J/\psi\rightarrow\eta\pi^0\phi$ reaction, let us fix the $K\bar K$ invariant mass to the  $\phi$ meson mass first. Neglecting the $K^*$ width, the triangle singularities for diagrams with the charged [Fig.~\ref{FIG1to3} (a) and (b)] and neutral [Fig.~\ref{FIG1to3} (c) and (d)] intermediate states are located at 1385.7~MeV and 1395.6~MeV, respectively.
The former is slightly above the $K^{*+}K^-$ threshold at 1385.3~MeV (we take the central values in Ref.~\cite{Tanabashi:2018oca} for the masses), and the latter is 2~MeV above the $K^{*0}\bar K^0$ threshold at 1393.6~MeV.
Therefore, we will see two sharp peaks in the $\pi^0\phi$ invariant mass distribution,\footnote{In this case, the singularities are at the physical boundary, which means that the triangle loop amplitudes have logarithmic singularities in the physical region.
This of course will not happen in the real physical case because for all the particles being on shell, the $K^*$ must be able to decay so that the singularities move to the complex plane.}, and each of the peak has a cusp at the $K^* \bar K$ threshold on its left shoulder. When the $K^*$ width is taken into account, the sharp peaks will be smeared to a smooth and much broader peak with a width dictated by the $K^*$ width. For a detailed study of the width effects on triangle singularity, we refer to Ref.~\cite{Debastiani:2018xoi,Du:2019idk}.

\begin{figure}
     \centering
     \includegraphics[width=0.5\textwidth]{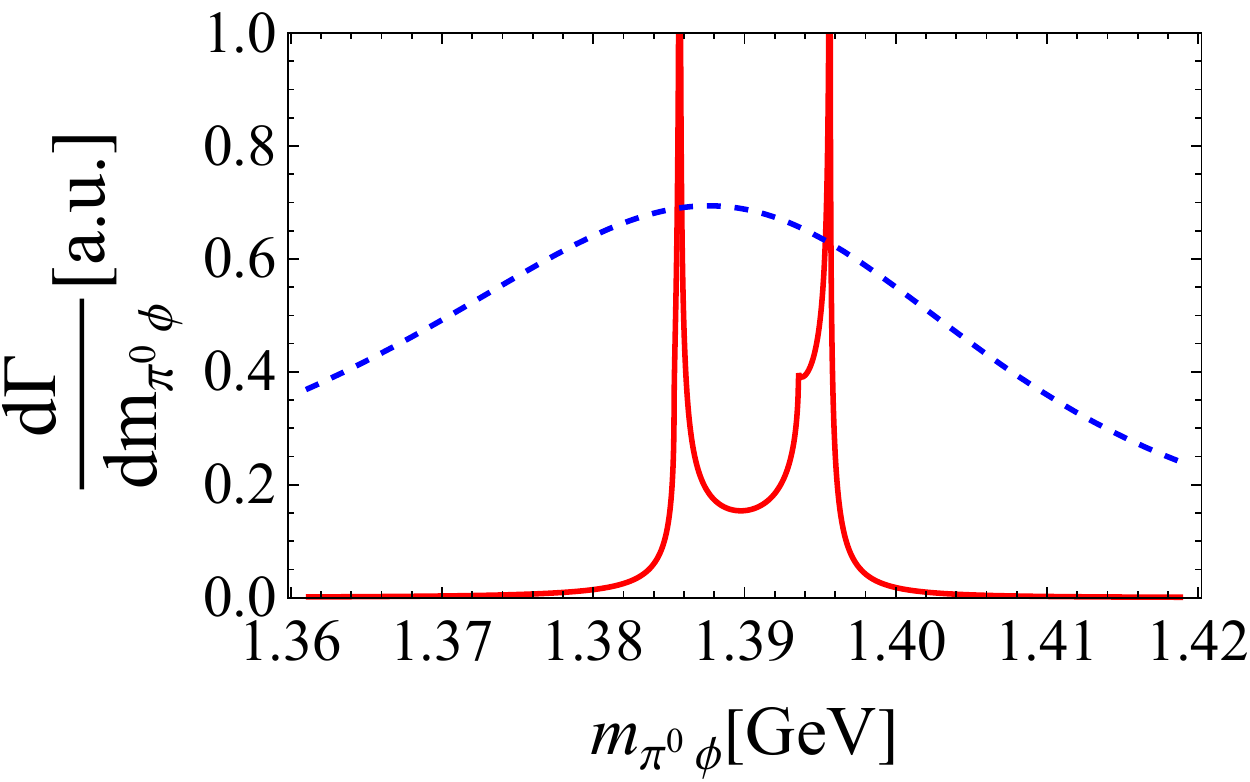}
     \caption{The $\pi^0\phi$ invariant mass distribution in arbitrary units (a.u.) for the decay $J/\psi \rightarrow \eta \pi^0 \phi$ from the mechanism shown in Fig.~\ref{FIG1to3}.
    The red-solid line shows $\der\Gamma_{J/\psi\to\eta\pi^0\phi}/{\der m}_{\pi^0\phi}$ in Eq.~\eqref{eq:dgamdm} obtained by switching off the $K^*$ width, and the blue-dashed curves corresponds to the result including the width of $K^*$.
     The $\pi^0\phi$ distribution with the $K^*$ width is multiplied by a factor of 500 in order to have a size comparable to the red-solid line.}
     \label{Resultsof1to3}
\end{figure}

In Fig.~\ref{Resultsof1to3}, we show the $\pi^0\phi$ invariant mass distribution given by Eq.~\eqref{eq:dgamdm} with and without considering the $K^*$ width.
The behavior of the curves is exactly as expected in the above.
The red-solid line is the ${\pi^0\phi}$ invariant mass distribution of $J/\psi\to\eta\pi^0\phi$ without the $K^*$ width.
Two singularities emerging from the charged and neutral $K^*\bar{K}K$ loops show up at around $1386$~MeV and $1396$~MeV, respectively\footnote{The structure at 1393.6~MeV is the $K^{*0}\bar K^0$ threshold cusp, and the $K^{*+}K^-$ threshold cusp is invisible in the plot because it is too close to (only 0.4~MeV below) the triangle singularity.}.
The dashed curve in the figure is obtained by using a complex mass $m_{K^*}-i\Gamma_{K^*}/2$ for the $K^*$ with $\Gamma_{K^*}=50$~MeV.
With the formula for the triangle singularity in Eq.~\eqref{eq:ts},
the two triangle singularities for the charged and neutral intermediate states are now located at $1385.6-i25.5$~MeV and $1395.5-i26.1$~MeV where the imaginary parts are introduced by the finite width of $K^*$. Since the difference between the two real parts is much smaller than twice the imaginary part, the two singularities lead to only a single broad peak shown as the dashed curve.
Thus, the two peaks from the charged and neutral $K^*\bar{K}K$ loops with isospin mass splitting turn into a single peak in the $\pi^0\phi$ distribution with the inclusion of the $K^*$ width.

The BESIII Collaboration reported the Dalitz plot distribution for the decay $J/\psi\rightarrow\eta\pi^0\phi$ in the $(m_{\eta\pi^0}^2,m_{\pi^0\phi}^2)$ plane~\cite{Ablikim:2018pik}.
Here, let us discuss the triangle singularity contribution to the Dalitz plot.
Note that the $a_0(980)$/$f_0(980)$ resonances seen in the Dalitz plot (along the $\pi^0\eta$ direction) in Ref.~\cite{Ablikim:2018pik}, as well as other resonances listed in the Review of Particle Physics~\cite{Tanabashi:2018oca}, are not considered here, because they cannot produce any nontrivial narrow structure in the $\pi^0\phi$ invariant mass distribution (as can be anticipated from the Dalitz plot projection), which is the focus here.
The mechanism discussed here is only important in a small energy region for $m_{\pi^0\phi}$ not far from the $K^*\bar K$ threshold. This means that the extension of our calculation to the whole phase space would not be adequate.
However, as we discuss below the main feature of the triangle singularity contribution to the Dalitz plot is consistent with what was observed by the BESIII Collaboration.

\begin{figure}
    \centering
    \includegraphics[width=0.5\textwidth]{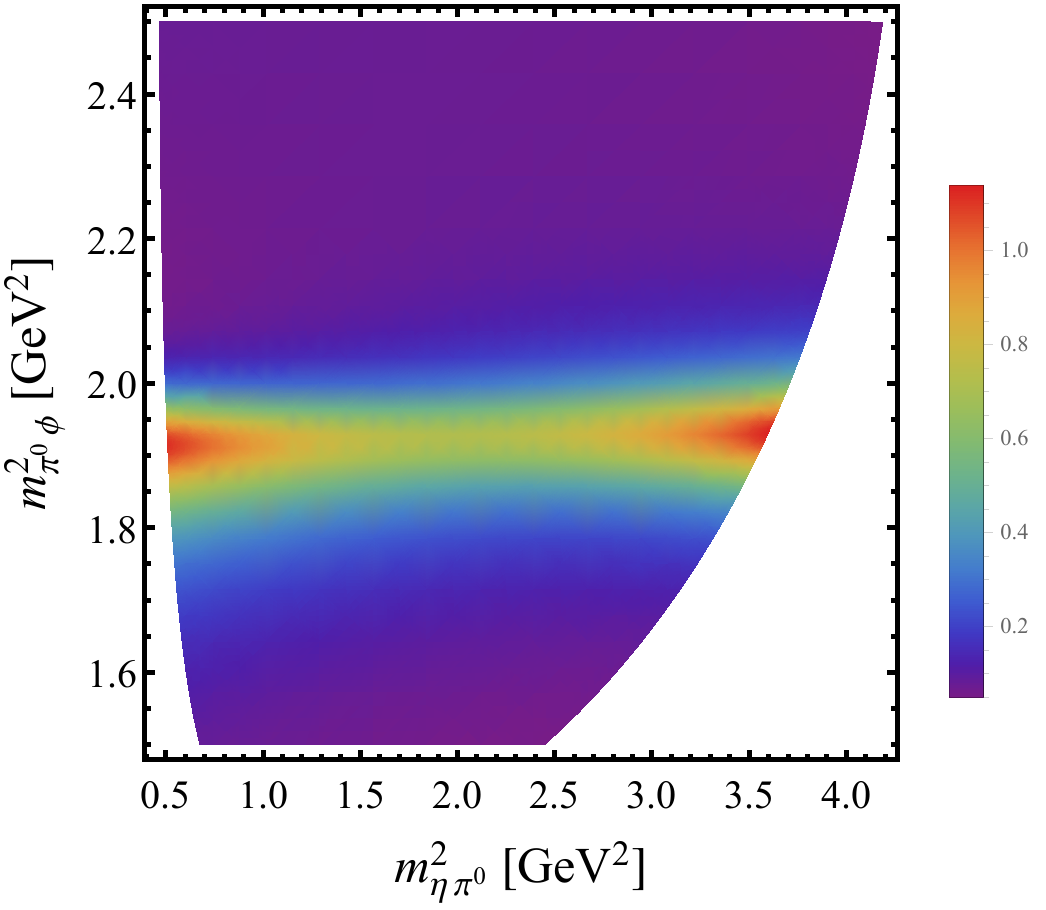}
    \caption{Dalitz plot of the $J/\psi\rightarrow\eta\pi^0\phi$ in the  $(m_{\eta\pi^0}^2,m_{\pi^0\phi}^2)$ plane
    from the triangle loop amplitude given in Eq.~\eqref{eq:mtot}.
    }
    \label{fig:dalitz}
\end{figure}
With Eq.~\eqref{eq:dgamdm}, the Dalitz plot distribution is given by
\begin{align}
    \frac{\der^2\Gamma}{\der m_{\eta\pi^0}^2\der m_{\pi^0\phi}^2}=\frac{1}{2|{\mathbf p}_\eta||{\mathbf p}_{\pi^0}|}\frac{\der^2\Gamma}{\der m_{\pi^0\phi}^2\der\cos\theta}\propto|\mM_{J/\psi\rightarrow\eta\pi^0\phi}|^2
\end{align}
with the amplitude $\mM_{J/\psi\rightarrow\eta\pi^0\phi}$ given in Eq.~\eqref{eq:mtot}.
Considering the $K^*$ width, the resulting Dalitz plot is shown in Fig.~\ref{fig:dalitz}. One can see that the peak shown as the dashed line in Fig.~\ref{Resultsof1to3} in fact shows up as an accumulation of events at both ends of the phase-space allowed region for $m_{\eta\pi^0}$.
The reason for such a behavior is that triangle singularity happens when all the particles move collinearly (see, e.g., discussions in Ref.~\cite{Bayar:2016ftu}), which corresponds to the boundary of the Dalitz plot.
Therefore, the effects should show up most prominently at $m_{\pi^0\phi}\sim 1.39$~GeV and at the two ends of the physical $m_{\pi^0\eta}$ region.
In the Dalitz plot given by the BESIII Collaboration with the $\eta$ reconstructed from two photons~\cite{Ablikim:2018pik}, there is a clear band at $m_{\pi^0\phi}\sim 1.4$~GeV, and the events along the band accumulate around the two ends. Except that in the low $m_{\pi^0\eta}$ region there is a large contribution from the $f_0(980)$ and $a_0(980)$ resonances, which are not considered here, the gross feature of the band is consistent with what is shown in Fig.~\ref{fig:dalitz}.
Data with higher statistics are called for in order to make a firm conclusion on whether the band is due to the triangle singularities discussed here or due to a resonance.

The loop integrals in the decay amplitudes are ultraviolet (UV) divergent, and we use dimensional regularization with the $\overline{\text{MS}}$ subtraction scheme to regularize the UV divergence (see Appendix~\ref{app:mM}). It is worthwhile to mention that triangle singularity happens when all the three intermediate particles go on shell and thus is an infrared singularity.
Thus, the UV divergence does not affect the presence of triangle singularity, but requires introducing a counterterm to absorb the  divergence.
Here, since we do not intend to construct a full model for the decay and focus only in the triangle singularity effects, we refrain from introducing the counterterm and simply take the dimensional regularization scale to be $\mu=1.2$~GeV in the calculation  of the Dalitz plot.

\begin{figure}[t]
    \centering
    \includegraphics[width=\textwidth]{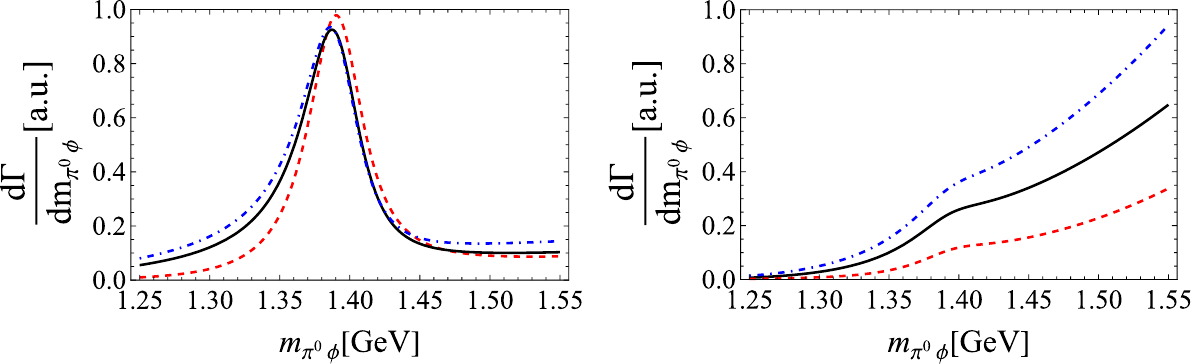}
    \caption{Dependence of the $\pi^0\phi$ distribution of $J/\psi\rightarrow\eta\pi^0\phi$ and $J/\psi\rightarrow\pi^0\pi^0\phi$ on the scale ($\mu$) in dimensional regularization with the $\overline{\text{MS}}$ subtraction scheme used in evaluating the loop integrals. The red dashed, black solid and blue dot-dashed lines correspond to $\mu = $ 0.2, 1.2 and 2.2~GeV, respectively.}
    \label{fig:mu-dep}
\end{figure}
The $m_{\pi^0\phi}$ distribution for the $J/\psi\rightarrow\eta\pi^0\phi$ including the $K^*$ width is shown in the left panel of Fig.~\ref{fig:mu-dep},
which is the same as  Fig.~\ref{Resultsof1to3} with the $K^*$ width up to an overall normalization factor. One can see a clear peak around $1.39$~GeV with a width about 50~MeV.
Because of the isospin breaking, the charged and neutral loops largely cancel each other in the region outside the peak, making the peak a prominent structure.
The dependence on the scale $\mu$ which appears in the regularization of the UV divergence is also checked:
the results with $\mu=0.2$, $1.2$, and $2.2$~GeV for $J/\psi\rightarrow\eta\pi^0\phi$ are shown as dashed, solid and dot-dashed lines, respectively, in the left panel of Fig.~\ref{fig:mu-dep}.
As one can see, the peak in the case of $J/\psi\rightarrow\eta\pi^0\phi$ is affected little by changing $\mu$.
This is because the decay breaks isospin, and the UV divergence part gets largely canceled in the difference between the contributions of the charged and neutral meson loops, see Eq.~\eqref{eq:mtot}.
Such a cancellation has been discussed in Ref.~\cite{Guo:2010ak} studying the charmed-meson loop contribution to the isospin breaking decay $\psi'\to J/\psi \pi^0$: it is shown that at the leading order of nonrelativistic effective field theory, the UV divergence in the neutral and charge meson loops cancel with each other, leaving a finite piece depending on the masses of intermediate particles.

\subsection{\texorpdfstring{\bm{$J/\psi \rightarrow \pi^0 \pi^0 \phi $}}{Jppipiphi}}

\label{sec:pipiphi}

Differently, the $J/\psi\rightarrow\pi^0\pi^0\phi$ reaction is isospin-symmetry allowed, and the charged and neutral loops add up to give the final result. Thus, one would expect the peak in the $\pi^0\phi$ invariant mass distribution due to triangle singularities to be much more modest than that in the $J/\psi\rightarrow\pi^0\eta\phi$ case, and the result should have a large scale dependence. This is indeed the case as can be seen from the right panel of Fig.~\ref{fig:mu-dep}. Nonetheless, the triangle singularities in this process have the same origin as the prominent one for the $J/\psi\rightarrow\pi^0\eta\phi$, and deserve to be studied in more detail to reveal another aspect of the $J/\psi\rightarrow\pi^0\pi^0\phi$ process in addition to the $\pi\pi$ distribution that has been extensively studied both experimentally~\cite{Ablikim:2004wn} and theoretically~\cite{Meissner:2000bc,Roca:2004uc,Lahde:2006wr,Liu:2009ub}.

\subsection{\texorpdfstring{\bm{$ J/\psi \rightarrow \eta \pi^0 K \bar{K} $}}{JpepKKbar}}

\begin{figure}[tb]
    \centering
    \includegraphics[width=\textwidth]{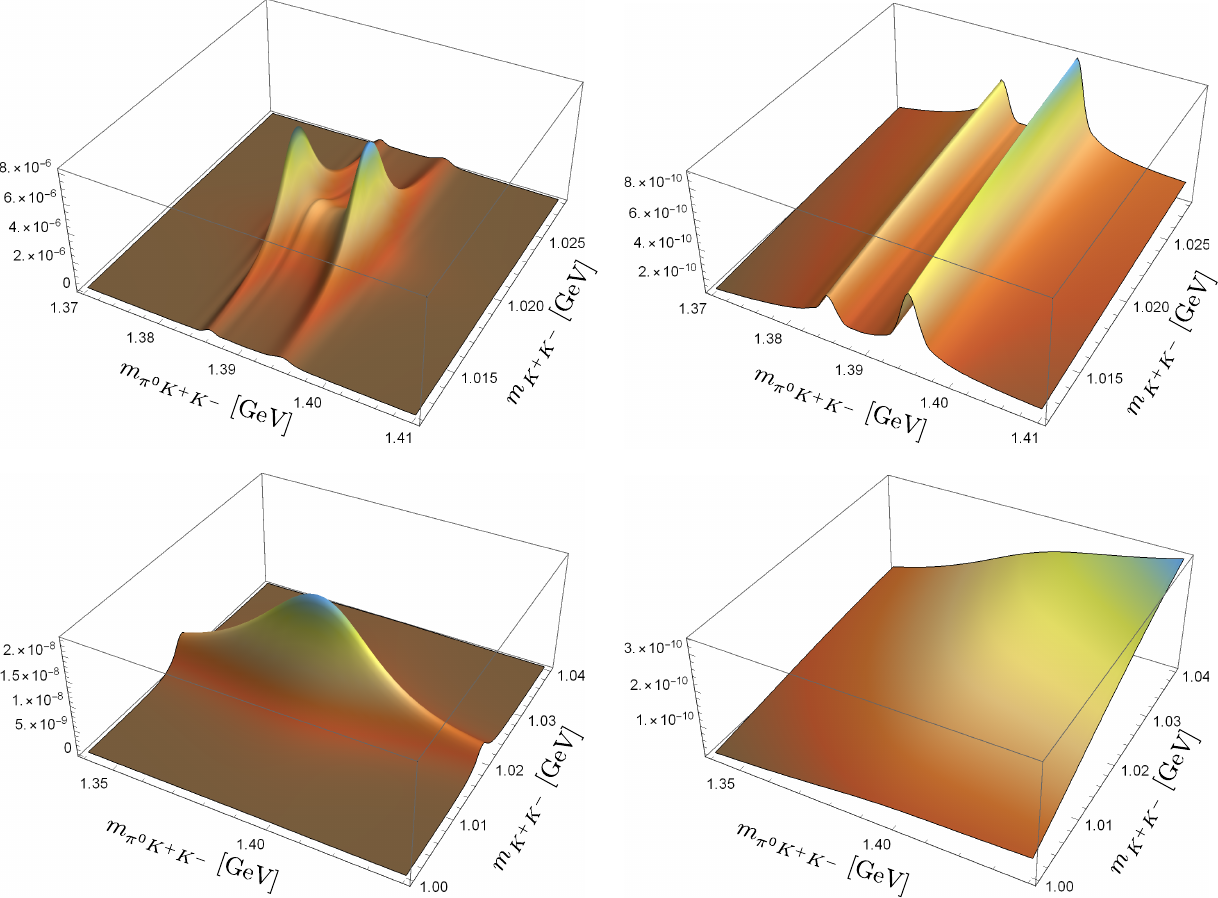}
    \caption{Differential width for the $J/\psi \rightarrow \eta \pi^0 K \bar{K}$ as a function of invariant masses of $K\bar K$ and $\pi^0 K \bar K$ in arbitrary units, by considering only triangle loop diagrams shown in Fig.~\ref{FIG1to4}.  The left (right) two plots show results considering triangle diagrams with $I=0$ ($I=1$) $K\bar{K}$ FSI.
    The $K^*$ width is considered in the lower-line plots, but not in the upper-line ones. }
    \label{Resultsof1to4Eta}
\end{figure}

One important feature of triangle singularity is that its effect is very sensitive to the kinematical variables such as masses of the intermediate particles and external energies.\footnote{It was recently proposed that this sensitivity can be used to make a very precise determination of the binding energy of the $X(3872)$ by measuring the line shape of the $X(3872)\gamma$ produced from a short-distance $D^{*0}\bar D^{*0}$ source~\cite{Guo:2019qcn}.} Using Eq.~\eqref{eq:range}, the triangle singularity is at the physical boundary (neglecting the $K^*$ width) only when the $K\bar K$ invariant mass is in the range of $[987.4, 1025.9]$~MeV and $[995.2, 1033.7]$~MeV for the charged and neutral intermediate states, respectively. The $\phi$ mass is just right in the range. When the $K\bar K$ invariant mass is pushing away from this range, the effects caused by the singularities will damp quickly.

Therefore, in order to reveal the origin of the band in the Dalitz plot at  $m_{\pi^0\phi}\sim1.4$~GeV in the BESIII data~\cite{Ablikim:2018pik}, i.e., whether it is indeed due to triangle singularities, we can study the $J/\psi\rightarrow\eta\pi^0K\bar{K}$ reaction and investigate the correlation between the $K\bar K$ invariant mass and the triangle singularity effects.
In Fig.~\ref{Resultsof1to4Eta}, we show the $(m_{\pi^0 K {\bar K}},m_{K {\bar K}})$ distribution of the $J/\psi\rightarrow\eta\pi^0K\bar{K}$ reaction considering only the loop diagrams shown in Fig.~\ref{FIG1to4} with (lower line) and without (upper line) considering the $K^*$ width. The upper two plots show clearly the two triangle singularities in the charge and neutral loops.\footnote{Notice that there should always be $K^{*+}K^-$ and $K^{*0}\bar K^0$ threshold cusps, which get smeared by the $K^*$ width and would produce a mild bump if there is no enhancement due to nearby triangle singularities or due to possible nearby resonances.}
The structures get smeared by the $K^*$ width in the lower two plots, while the one with $I=0$ $K\bar K$ still shows a clear peak at $m_{\pi^0K^+K^-}\sim1.4$~GeV due to isospin breaking, which is most prominent at $m_{K^+K^-}\simeq1.02$~GeV due to the presence of the $\phi$ resonance.
When the $K\bar K$ invariant mass is away from the $\phi$ resonant region, the peak along  $m_{\pi^0K^+K^-}$ also becomes much less evident in the lower left plot.
Were the band at $m_{\pi^0\phi}\sim1.4$~GeV in the Dalitz plot observed by the BESIII Collaboration due to a resonance decaying into $\pi^0\phi$, the same resonance would also be able to decay into $\pi^0 K^+ K^-$, and its signal should always be there no matter what value the $K^+K^-$ invariant mass takes (as long as there are enough data).

\begin{figure}[tb]
    \centering
    \includegraphics[width=0.45\textwidth]{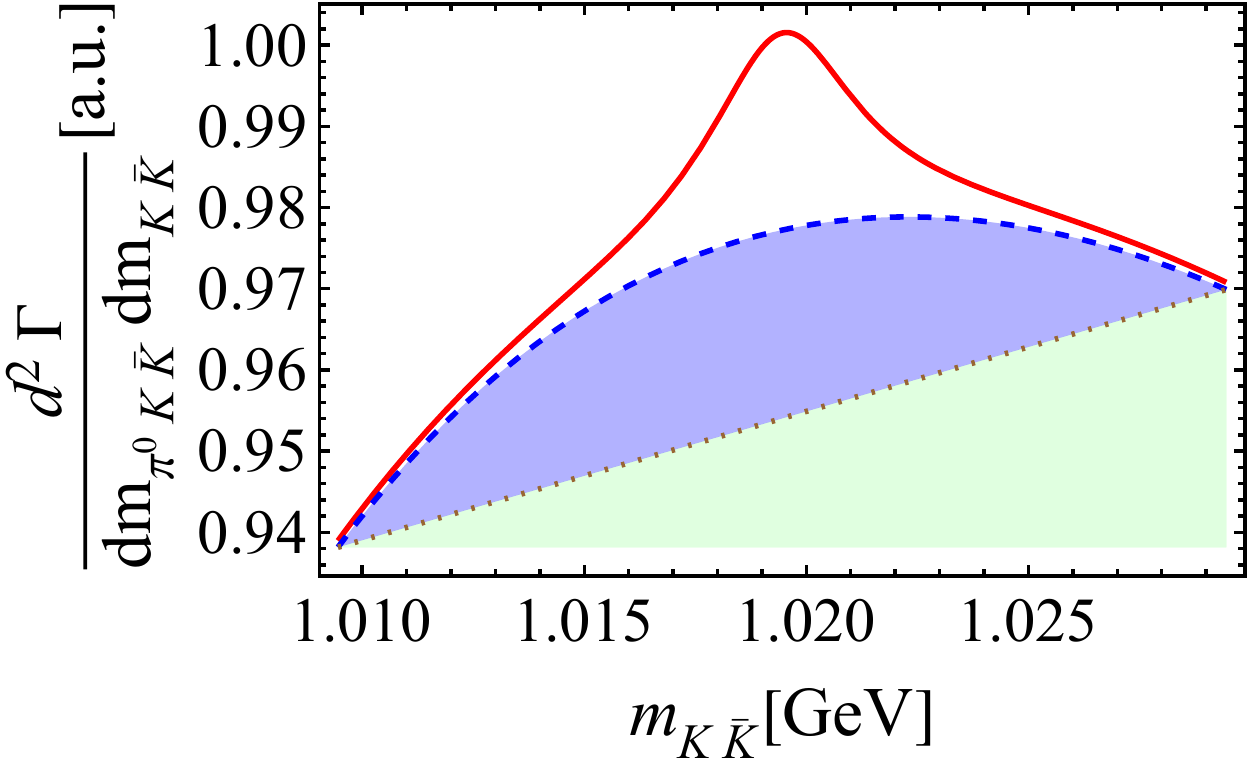}
    \includegraphics[width=0.45\textwidth]{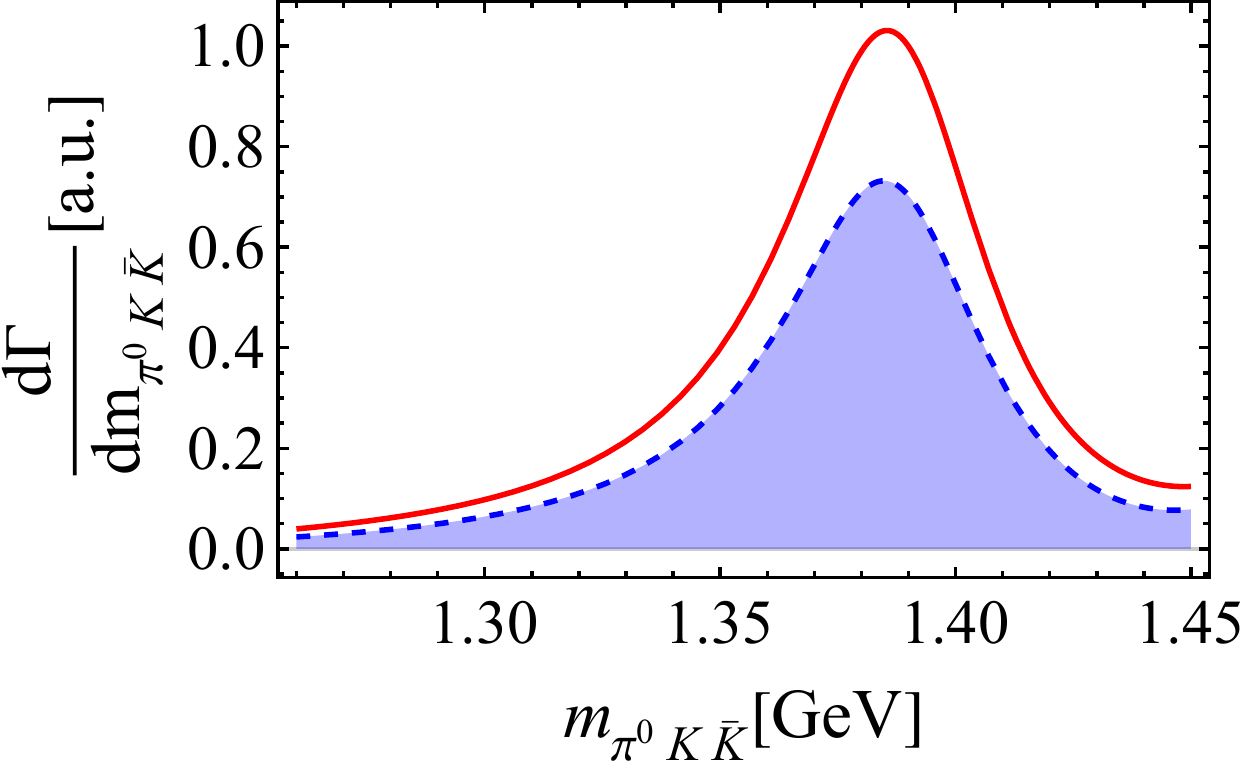}
    \caption{Left panel: the $K^+K^-$ invariant mass distribution in the $\phi$ window  with $m_{\pi^0 K \bar{K}}$ fixed at $1.39$ GeV for $J/\psi \rightarrow \eta \pi^0 K \bar{K}$. The solid line includes all the diagrams (tree, loops with the $I=0$ and $I=1$ $K\bar K$ FSI) in Fig.~\ref{FIG1to4}, the dashed line includes the tree and $I=0$ $K\bar K$ FSI diagrams, and the dotted line is a linear interpolation between the two ends of a narrow window of the $\phi$ resonance: $m_{K\bar K}\in[1010,1030]$~MeV. Right panel: the $\pi^0K^+K^-$ invariant mass distribution of the same reaction with $m_{K\bar K}$ integrated in the $\phi$ window, where we have subtracted the sideband contribution (i.e., the part below the straight dotted line in the left panel). The $K^*$ width has been taken into account in these plots.
    }
\label{ResultsOf1to4EtaInt}
\end{figure}

In the left panel of Fig.~\ref{ResultsOf1to4EtaInt}, we show the $m_{K {\bar K}}$ distribution at $m_{\pi^0 K {\bar K}}=1.39$~GeV in a window of the $K\bar K$ invariant mass containing the $\phi$ resonance, i.e., $m_{K\bar K}\in[1010,1030]$~MeV as that taken in the BESIII paper~\cite{Ablikim:2018pik}, given by Eq.~\eqref{dGammaOf1to4Eta} with both tree and loop amplitudes in Eq.~\eqref{AmpOf1to4Eta} with the $K^*$ width. One clearly sees the $\phi$ peak. In the BESIII measurement, the $\phi$ events are selected in a $K^+K^-$ invariant mass window of  $[m_\phi-10~\text{MeV}, m_\phi+10~\text{MeV}]$, and a sideband subtraction is made so as to remove the background events  that are not from the $\phi$. Here, similarly we make a linear interpolation between the two ends of the $\phi$ window to take the sideband subtraction, see the caption of Fig.~\ref{ResultsOf1to4EtaInt}. The right panel shows the $\pi^0 K^+K^-$ invariant mass distribution  by integrating  Eq.~\eqref{dGammaOf1to4Eta} over $m_{K \bar{K}}$ in $\phi$ window, and the sideband subtraction has been made. A clear peak shows up at around 1.39~GeV with little background.

At last, we make a brief comment on the Schmid theorem~\cite{Schmid:1967ojm} that can matter when the external and internal particles are the same in the triangle diagram (i.e., when the rescattering of particles $m_2$ and $m_3$ in Fig.~\ref{fig:triangle} is elastic).
The Schmid theorem claims that the triangle singularity contribution does not show up in the invariant mass distribution which is obtained by projecting the Dalitz plot into one invariant mass.
The validity and limitations of the Schmid theorem has been studied in Refs.~\cite{Aitchison:1969tq,Anisovich:1995ab,Szczepaniak:2015hya,Debastiani:2018xoi}. It is pointed out that the theorem holds only for the single-channel case, and in the case of very small width of the intermediate particles such that the triangle diagram can be well approximated by solely the singularity contribution.
In such a situation, the sum of the tree-level and triangle diagram is approximately given by the tree-level amplitude multiplied by a phase factor with the phase given by the scattering phase shift of the two particles that rescatter. As a result, the invariant mass distribution does not have a sharp peak, but behaves as that of the tree-level $t$-channel exchange diagram.
One sees that the Schmid theorem is not applicable here for two reasons: (1) the $K^*$ meson has a width of about 50~MeV; (2) for the reaction of interest $J/\psi\to \eta\pi^0K^+K^-$, only the $K^+K^-$ pair appears in the final state while both the $K^+K^-$ and $K^0\bar{K}^0$ appear in the loops.

\section{Summary}
\label{sec:summary}

In summary, by considering the triangle diagrams with $K^*K\bar K$ intermediate particles for the isospin breaking process $J/\psi \rightarrow \eta \pi^0 \phi$ shown in Fig.~\ref{FIG1to3}, we show that there appears a band in the Dalitz plot which is located at the $\pi^0 \phi$ invariant mass of around 1.4~GeV by virtue of triangle singularities.
Events along the band accumulate around the two ends close to the Dalitz plot boundary. Such a feature is consistent with what was observed by the BESIII Collaboration with the $\eta$ reconstructed from two photons in Ref.~\cite{Ablikim:2018pik}.

In order to check whether this is the genuine mechanism behind the experimental observation, we further study the processes $J/\psi \to \pi^0\pi^0\phi$ and $J/\psi\to \eta\pi^0 K^+K^-$. These processes are dominated by the isospin conserving contribution. It is shown that the triangle singularity mechanism has a much more modest effect for the  $J/\psi \to \pi^0\pi^0\phi$ and leads to a broader bump. As for the $J/\psi\to \eta\pi^0 K^+K^-$, our results show that the triangle singularity effect is most evident with the $K^+K^-$ invariant mass is in the $\phi$ resonance mass region where isospin breaking plays an important role. Thus, by taking events with $m_{K^+K^-}$ away from that region, we expect the band at $m_{\pi^0 K^+K^-}\sim 1.4$~GeV in the Dalitz plot would become weaker and eventually invisible if $K^+K^-$ is sufficiently away from the $\phi$ mass region. On the other hand, if the band in the Dalitz plot is due to a resonance, which should also be able to decay into $\pi^0K^+K^-$, one expects the band remains independent of the $K^+K^-$ invariant mass.

Therefore, in order to reveal the physics behind the band at $m_{\pi^0\phi}\sim1.4$~GeV observed by the BESIII Collaboration, we suggest experimentalists to take more data for the $J/\psi \rightarrow \eta\pi^0\phi$ to make the Dalitz plot more clear, and to check whether the structure persists in other $K^+K^-$ invariant mass regions.

%%%%%%%%%%%%%%%%%%%%%%%%%%%%%%%%%%%
\begin{acknowledgements}
We are grateful to Eulogio Oset for useful comments. This work is supported in part by the National Natural Science Foundation of China (NSFC) and  the Deutsche Forschungsgemeinschaft (DFG) through the funds provided to the Sino-German Collaborative Research Center ``Symmetries and the Emergence of Structure in QCD"  (NSFC Grant No. 11621131001, DFG Grant No. TRR110), by the NSFC under Grant No. 11847612 and No. 11835015, by the Chinese Academy of Sciences (CAS) under Grant No. QYZDB-SSW-SYS013 and No. XDPB09, and by
the CAS Center for Excellence in Particle Physics (CCEPP). S.S. is also supported by the 2019 International Postdoctoral Exchange Progarm and the CAS President’s International Fellowship Initiative (PIFI) under Grant No.~2019PM0108.
\end{acknowledgements}

\appendix
\section{Explicit form of \texorpdfstring{\bm{$\mM_{id}^{\mu\nu}$}}{Mmunu}\label{app:mM}}

In this section we present the explicit form of $\mM_{id}^{\mu\nu}$ appearing in Eq.~\eqref{AmpOf1to3}.
First we define $I_{id}$, $J_{id}$, and $I_{id}^{\mu_1\mu_2...\mu_n}$ as
\begin{align}
    &I_{id}
    =\int\frac{d^4q}{(4\pi)^2}\frac{1}{[q^2-m_{K_{id}}^2+i\epsilon][(q+k_1)^2-m_{K^*_{id}}^2+i\epsilon][(q+k_2)^2-m_{K_{id}}^2+i\epsilon]},\\
    &J_{id}
    =\int\frac{d^4q}{(4\pi)^2}\frac{1}{[(q+k_1)^2-m_{K^*_{id}}^2+i\epsilon][(q+k_2)^2-m_{K_{id}}^2+i\epsilon]},\\
    &I_{id}^{\mu_1\mu_2...\mu_n}
    =\int\frac{d^4q}{(4\pi)^2}\frac{q^{\mu_1}q^{\mu_2}...q^{\mu_n}}{[q^2-m_{K_{id}}^2+i\epsilon][(q+k_1)^2-m_{K^*_{id}}^2+i\epsilon][(q+k_2)^2-m_{K_{id}}^2+i\epsilon]}.
\end{align}
With the technique of tensor decomposition of loop integrals, $I^\mu$, $I^{\mu\nu}$, $I^{\mu\nu\rho}$, and $I^{\mu\nu\rho\sigma}$ can be rewritten as given in Ref.~\cite{Hahn:1998yk},
\begin{align}
    I^\mu=&\sum_{i=1,2}k_{i}^\mu C_i,\\
    I^{\mu\nu}=&g^{\mu\nu}C_{00}+\sum_{i,j=1,2}k_{i}^\mu k_{j}^\nu C_{ij},\\
    I^{\mu\nu\rho}=&\sum_{i=1,2}(g^{\mu\nu}k_i^\rho+g^{\nu\rho}k_i^\mu+g^{\rho\mu}k_i^\nu)C_{00i}+\sum_{i,j,k=1,2}k_{i}^\mu k_{j}^\nu k_{k}^\rho C_{ijk},\\
    I^{\mu\nu\rho\sigma}=&(g^{\mu\nu}g^{\rho\sigma}+g^{\mu\rho}g^{\nu\sigma}+g^{\mu\sigma}g^{\nu\rho})C_{0000} \notag \\
    &+\sum_{i,j=1,2}(g^{\mu\nu}k_i^\rho k_j^\sigma+g^{\nu\rho}k_i^\mu k_j^\sigma+g^{\rho\mu}k_i^\nu k_j^\sigma+g^{\mu\sigma}k_i^\nu k_j^\rho+g^{\nu\sigma}k_i^\mu k_j^\rho+g^{\rho\sigma}k_i^\mu k_j^\nu)C_{00ij} \notag \\
    &+\sum_{i,j,k,l=1,2}k_{i}^\mu k_{j}^\nu k_{k}^\rho k_{l}^\sigma C_{ijkl}.
\end{align}
Using these $I_{id}$ and $I^{\mu_1\mu_2...\mu_n}_{id}$, the loop amplitude $\mM_{id}^{\mu\nu}$ is written as
\begin{align}
\begin{split}
    \mM_{id}^{\mu\nu}=&I_{id}\left[k_1^\mu k_2^\nu-2k_2^\mu k_2^\nu+\frac{2(k_1\cdot k_2)}{m_{K_{id}^*}^2}k_1^\mu k_2^\nu-\frac{k_1^2}{m_{K_{id}^*}^2}k_1^\mu k_2^\nu\right]\\
    &+\frac{2}{m_{K_{id}^*}^2}I_{id}^\rho k_1^\mu k_2^\nu k_{2\rho}
    +\frac{1}{m_{K_{id}^*}^2}g_{\rho\sigma}I_{id}^{\rho\sigma}k_1^\mu k_2^\nu
    +I_{id}^\mu\left[-k_2^\nu+\frac{2(k_1\cdot k_2)}{m_{K_{id}^*}^2}k_2^\nu-\frac{k_1^2}{m_{K_{id}^*}^2}k_2^\nu\right]\\
    &+I_{id}^\nu\left[2k_1^\mu-4k_2^\mu+\frac{4(k_1\cdot k_2)}{m_{K_{id}^*}^2}k_1^\mu-\frac{2k_1^2}{m_{K_{id}^*}^2}k_1^\mu\right]
    +\frac{2}{m_{K_{id}^*}^2}I_{id}^{\mu\rho}k_2^\nu k_{2\rho}
    \\
    &+\frac{1}{m_{K_{id}^*}^2}g_{\rho\sigma}I_{id}^{\mu\rho\sigma}k_2^\nu+\frac{4}{m_{K_{id}^*}^2}I^{\nu\rho}k_1^\mu k_{2\rho}+\frac{2}{m_{K_{id}^*}^2}g_{\rho\sigma}I^{\nu\rho\sigma}k_1^\mu
    \\
    &+I_{id}^{\mu\nu}\left[\frac{4(k_1\cdot k_2)}{m_{K_{id}^*}^2}-2-\frac{2k_1^2}{m_{K_{id}^*}^2}\right]+\frac{4}{m_{K_{id}^*}^2}I_{id}^{\mu\nu\rho}k_{2\rho}+\frac{2}{m_{K_{id}^*}^2}g_{\rho\sigma}I_{id}^{\mu\nu\rho\sigma}.
\end{split}
\label{eq:tnsrdcmp}
\end{align}

Since both the $K^*K\pi^0$ and $\phi K \bar{K}$ vertices are $P$-wave and the $K^*$ propagator has a momentum dependent numerator, the amplitude of the triangle diagram is UV divergent. In all calculations of this paper, we use dimensional regularization and with the $\overline{\text{MS}}$ subtraction scheme for the subtraction of the UV divergence, and unless specified we take $\mu = 1.2$~GeV for the  scale introduced in dimensional regularization.

In Eq.~\eqref{eq:tnsrdcmp}, for example, a term $g_{\rho\sigma}I^{\rho\sigma}$ should be understood as
\begin{align}
g_{\rho\sigma}I_{id}^{\rho\sigma}=&\int\frac{d^4q}{(2\pi)^4}\frac{g_{\rho\sigma}q^\rho q^\sigma}{[q^2-m_{K_{id}}^2+i\epsilon][(q+k_1)^2-m_{K^*_{id}}^2+i\epsilon][(q+k_2)^2-m_{K_{id}}^2+i\epsilon]} \notag\\
=&\int\frac{d^4q}{(2\pi)^4}\left\{\frac{m_{K_{id}}^2}{[q^2-m_{K_{id}}^2+i\epsilon][(q+k_1)^2-m_{K^*_{id}}^2+i\epsilon][(q+k_2)^2-m_{K_{id}}^2+i\epsilon]}\right. \notag\\
&\left.\hspace{2cm}+\frac{1}{[(q+k_1)^2-m_{K^*_{id}}^2+i\epsilon][(q+k_2)^2-m_{K_{id}}^2+i\epsilon]}\right\} \notag\\
=&m_{K_{id}}^2I_{id}+J_{id}.
\end{align}
The terms $g_{\rho\sigma}I^{\mu\rho\sigma}$ and $g_{\rho\sigma}I^{\mu\nu\rho\sigma}$ also need similar manipulations.

\section{General phase space formula for \texorpdfstring{\bm{$1$}}{1} to \texorpdfstring{\bm{$n$}}{n}-body decay\label{app:PSV}}
The partial decay rate of a particle of mass $m$ into $n$ bodies in its rest frame is given in terms of the Lorentz-invariant matrix element ${\cal M}$ by
\begin{align}
\mathrm{d}\Gamma = \frac{(2\pi)^4} {2 m} |{\cal M}|^2 {\mathrm d}\Phi_n (p_1,\cdots,p_n),
\label{DecayFormulaPDG}
\end{align}
where ${\mathrm d}\Phi_n$ is an element of the $n$-body phase space given by
\begin{align}
{\mathrm d}\Phi_n (p_1,\cdots,p_n) = \delta^4\left(p-\sum_{i=1}^n p_i\right) \prod_{i=1}^n \frac{{\mathrm d}^3 p_i}{(2 \pi)^3 2 E_i} .
\label{PhaseSpace0}
\end{align}

To get the decay formula, we need to integrate the order-$4$ Dirac $\delta$-function in Eq.~\eqref{PhaseSpace0} which represents the four-momentum conservation. For instance, we can integrate the momenta of the last particle first:
\be
\begin{split}
&\int \frac{{\mathrm d}^3 p_n}{(2 \pi)^3 2 E_n} \delta^4\left(p-\sum_{i=1}^n p_i\right) = \int \frac{{\mathrm d}^4 p_n}{(2 \pi)^4} (2\pi) \delta\left(p^2_n-m^2_n\right) \theta(p^0_n) \delta^4\left(p-\sum_{i=1}^n p_i\right) \\
&\hspace{5mm}=\frac{1}{(2\pi)^3}\delta\left[\left(p-\sum_{i=1}^{n-1} p_i\right)^2-m^2_n\right] \theta\left[\left(p-\sum_{i=1}^{n-1} p_i\right)^0\right] \int {\mathrm d}^4 p_n \delta^4\left(p-\sum_{i=1}^n p_i\right) \\
&\hspace{5mm}\equiv\frac{1}{(2\pi)^3}\delta\left[\left(p-\sum_{i=1}^{n-1} p_i\right)^2-m^2_n\right] \theta\left[\left(p-\sum_{i=1}^{n-1} p_i\right)^0\right]\vartheta_n^4\left(p-\sum_{i=1}^n p_i\right),
\end{split}
\label{PhaseSpaceIntN}
\ee
where we have defined an order-$n$ $\vartheta$-function as follows:
\be
\int_X {\mathrm d}^n x \delta^n(x-x_0) \equiv \vartheta_X^n(x-x_0).
\ee
Substituting Eq.~\eqref{PhaseSpaceIntN} into Eq.~\eqref{PhaseSpace0}, we can get
\be
{\mathrm d}\Phi_n (p_1,\cdots,p_n) = \frac1{(2\pi)^3} \prod_{i=1}^{n-1} \frac{{\mathrm d}^3 p_i}{(2 \pi)^3 2 E_i} \delta\left[\left(p-\sum_{i=1}^{n-1} p_i\right)^2-m^2_n\right] \theta\left[\left(p-\sum_{i=1}^{n-1} p_i\right)^0\right] \vartheta_n^4\left(p-\sum_{i=1}^n p_i\right).
\label{dPhi(n-1)}
\ee
So far, we still have an order-1 Dirac $\delta$-function which represents the mass-shell condition of the last particle.
Just like what we did in Eq.~\eqref{PhaseSpaceIntN}, to eliminate the remaining order-1 Dirac $\delta$-function, we can integrate over the momentum of the $(n-1)$-th particle:
\be
\begin{split}
\int \frac{{\mathrm d}^3 p_{n-1}}{(2 \pi)^3 2 E_{n-1}}\delta\left[\left(p-\sum_{i=1}^{n-1} p_i\right)^2-m^2_n\right] &= \frac{1}{2(2\pi)^3} \int \frac{|\mathbf{p}_{n-1}|^2{\mathrm d}|\mathbf{p}_{n-1}|{\mathrm d}\Omega_{n-1}}{E_{n-1}} \delta\left[\left(p-\sum_{i=1}^{n-1} p_i\right)^2-m^2_n\right]\\
&=\frac{1}{2(2\pi)^3} \int |\mathbf{p}_{n-1}|{\mathrm d}E_{n-1}{\mathrm d}\Omega_{n-1} \delta\left[\left(p-\sum_{i=1}^{n-1} p_i\right)^2-m^2_n\right].
\end{split}
\label{PhaseSpaceInt(N-1)}
\ee
Before we continue to simplify the expression, let us make some definitions to facilitate subsequent calculations:
\be
\begin{split}
    p_{(k)} &= \sum^n_{i=k}p_i, \qquad \qquad
    p = \sum^n_{i=1}p_i = p_{(1)}, \notag \\
    m_{(k)} &= \sqrt{p^2_{(k)}},  \qquad \qquad
    E_{(k)} = p^0_{(k)} = \sqrt{\mathbf{p}_{(k)}^2 + m^2_{(k)}}.
\end{split}
\ee
In addition, we use $X^{*(l)}$ to represent the specific form of physical quantity $X$($X$ can be momentum $p$, energy $E$ or mass $m$ etc.) in the CM frame of the particle system $l,l+1,\cdots,n$.
Furthermore, for simplicity, we redefine $X^{*(l)}_l$ and $X^{*(l)}_{(l)}$ as  $X^{(*)}_l$ and $X^{*}_{(l)}$, respectively.
Since the element of final state phase space of each particle is Lorentz invariant, one can calculate each part in any reference frame.
Then, let us come back to the integration of the phase space for the $(n-1)$-th particle, and we calculate this part in the CM frame of $(n-1)$-th and $n$-th particle system:
\be
\begin{split}
    \int |\mathbf{p}_{n-1}|{\mathrm d}E_{n-1}{\mathrm d}\Omega_{n-1} \delta\left[\left(p-\sum_{i=1}^{n-1} p_i\right)^2-m^2_n\right] &= \int |\mathbf{p}_{n-1}|{\mathrm d}E_{n-1}{\mathrm d}\Omega_{n-1} \delta\left[(p_{(n-1)}-p_{n-1})^2-m^2_n\right]\\
    &=\int |\mathbf{p}^{(*)}_{n-1}|{\mathrm d}E^{(*)}_{n-1}{\mathrm d}\Omega^{(*)}_{n-1} \delta\left[(p^*_{(n-1)}-p^{(*)}_{n-1})^2-m^2_n\right].
\end{split}
\label{PhaseSpaceInt(N-1)star}
\ee
Here we note that in the CM frame of the $(n-1)$-th and $n$-th particle system we have
\be
    p^*_{(n-1)} = (m_{(n-1)},\mathbf{0})
    \label{pstar(n-1)}.
\ee
Substituting Eq.~\eqref{pstar(n-1)} into the order-1 Dirac $\delta$-function of Eq.~\eqref{PhaseSpaceInt(N-1)star},
\be
\begin{split}
    \delta\left[(p^*_{(n-1)}-p^{(*)}_{n-1})^2-m^2_n\right] &= \delta(m^2_{(n-1)}+m^2_{n-1}-2m_{(n-1)}E^{(*)}_{n-1}-m^2_n) \\
    &= \frac{1}{2m_{(n-1)}}\delta\left(E^{(*)}_{n-1}-\frac{m^2_{(n-1)}+m^2_{n-1}-m^2_n}{2m_{(n-1)}}\right).
\end{split}
\label{nthParticleOnshellCondition}
\ee

Then, using Eq.~\eqref{nthParticleOnshellCondition}, Eq.~\eqref{PhaseSpaceInt(N-1)star} is written as
\be
\begin{split}
    \int |\mathbf{p}^{(*)}_{n-1}|{\mathrm d}E^{(*)}_{n-1}{\mathrm d}\Omega^{(*)}_{n-1} \delta&\left[(p^*_{(n-1)}-p^{(*)}_{n-1})^2-m^2_n\right] \\
     & = \int |\mathbf{p}^{(*)}_{n-1}|{\mathrm d}\Omega^{(*)}_{n-1} \frac{1}{2m_{(n-1)}} \vartheta_{n-1}\left(E^{(*)}_{n-1}-\frac{m^2_{(n-1)}+m^2_{n-1}-m^2_n}{2m_{(n-1)}}\right),
\end{split}
\label{(n-1)thInt}
\ee
where we have integrated out the last order-1 Dirac $\delta$-function.
Then substituting Eq.~\eqref{(n-1)thInt} into Eq.~\eqref{PhaseSpaceInt(N-1)}, we can get:
\be
\begin{split}
    {\mathrm d}\Phi_n (p_1,\cdots,p_n) =&\frac1{2(2\pi)^6} \left(\prod_{i=1}^{n-2} \frac{{\mathrm d}^3 p_i}{(2 \pi)^3 2 E_i}\right)\frac{|\mathbf{p}^{(*)}_{n-1}|{\mathrm d}\Omega^{(*)}_{n-1}}{2m_{(n-1)}}\theta\left[(p-\sum_{i=1}^{n-1} p_i)^0\right] \\
    &\vartheta_n^4\left(p-\sum_{i=1}^n p_i\right)\vartheta_{n-1}\left(E^{(*)}_{n-1}-\frac{m^2_{(n-1)}+m^2_{n-1}-m^2_n}{2m_{(n-1)}}\right).
\end{split}
\ee
It is worth noting that the effect of the $\theta$ and $\vartheta$ functions is to limit the integrating range to the physical region, and it is easy to check that in the physical region,
\be
\theta\left[(p-\sum_{i=1}^{n-1} p_i)^0\right] \vartheta_n^4\left(p-\sum_{i=1}^n p_i\right)\vartheta_{n-1}\left(E^{(*)}_{n-1}-\frac{m^2_{(n-1)}+m^2_{n-1}-m^2_n}{2m_{(n-1)}}\right) = 1.
\ee
Thus, by confining the integral range to the physical region, we can get
\be
\begin{split}
    {\mathrm d}\Phi_n (p_1,\cdots,p_n) &= \frac1{2(2\pi)^6}\left(\prod_{i=1}^{n-2} \frac{{\mathrm d}^3 p_i}{(2 \pi)^3 2 E_i}\right)\frac{|\mathbf{p}^{(*)}_{n-1}|{\mathrm d}\Omega^{(*)}_{n-1}}{2m_{(n-1)}} \\
    &= \frac1{2^n(2\pi)^{3n}}\left(\prod_{i=1}^{n-2}|\mathbf{p}_i|{\mathrm d}E_i{\mathrm d}\Omega_i\right)\frac{|\mathbf{p}^{(*)}_{n-1}|{\mathrm d}\Omega^{(*)}_{n-1}}{m_{(n-1)}}.
\end{split}
\ee

This expression is sufficient to get the result, but it is better to express it in terms of invariant masses.
Then, let us consider the relation between energy and invariant masses for each particle:
\be
\begin{split}
    E^{(*)}_k &=\sqrt{\mathbf{p}^{{(*)}2}_k+m^2_k}=\sqrt{\mathbf{p}^{*(k)2}_{(k+1)}+m^2_k}=\sqrt{E^{*(k)2}_{(k+1)}-m^2_{(k+1)}+m^2_k}, \\
    E^{(*)}_k &=m_{(k)}-E^{*(k)}_{(k+1)}.
\end{split}
\ee
With these two equations, we can get
\be
    m_{(k)}\mathrm{d}E^{(*)}_k = (m_{(k)}-E^{(*)}_k)\mathrm{d}m_{(k)}-m_{(k+1)}\mathrm{d}m_{(k+1)}.
    \label{dmkchange}
\ee
With $k=1$, because of $\mathrm{d}m_{(1)}=\mathrm{d}m=0$,
we have
\be
    m\mathrm{d}E^{(*)}_1 = -m_{(2)}\mathrm{d}m_{(2)},
    \label{dE1anddm(2)}
\ee
and we obtain
\be
\begin{split}
    & {\mathrm d}\Phi_n (p_1,\cdots,p_n) \\ =&\,\frac1{2^n(2\pi)^{3n}}\left(\prod_{i=1}^{n-2}|\mathbf{p}^{(*)}_i|{\mathrm d}E^{(*)}_i{\mathrm d}\Omega^{(*)}_i\right)\frac{|\mathbf{p}^{(*)}_{n-1}|{\mathrm d}\Omega^{(*)}_{n-1}}{m_{(n-1)}} \notag \\
    =& \frac1{2^n(2\pi)^{3n}}\left(\prod_{i=1}^{n-2}|\mathbf{p}^{(*)}_i|{\mathrm d}\Omega^{(*)}_i\right)\frac{|\mathbf{p}^{(*)}_{n-1}|{\mathrm d}\Omega^{(*)}_{n-1}}{m_{(n-1)}}\left(\prod_{k=1}^{n-2}{\mathrm d}E^{(*)}_k\right)  \\
    =& \frac1{2^n(2\pi)^{3n}}\left(\prod_{i=1}^{n-2}|\mathbf{p}^{(*)}_i|{\mathrm d}\Omega^{(*)}_i\right)\frac{|\mathbf{p}^{(*)}_{n-1}|{\mathrm d}\Omega^{(*)}_{n-1}}{m_{(n-1)}} \left|\bigwedge_{k=1}^{n-2}\frac{ (m_{(k)}-E^{(*)}_k)\mathrm{d}m_{(k)}-m_{(k+1)}\mathrm{d}m_{(k+1)}}{ m_{(k)}}\right|,
\end{split}
\ee
where $\bigwedge$ represents the wedge product which has the following properties,
\be
    \mathrm{d}x\bigwedge\mathrm{d}y=-\mathrm{d}y\bigwedge\mathrm{d}x , \qquad \mathrm{d}x\bigwedge\mathrm{d}x=0.
\ee
With Eq.~\eqref{dE1anddm(2)}, the differential invariant mass $\mathrm{d}m_{(k)}$ in $\mathrm{d}E^*_k$ vanishes, and thus we have
\begin{align}
     {\mathrm d}\Phi_n (p_1,\cdots,p_n) &=\frac1{2^n(2\pi)^{3n}}\prod_{i=1}^{n-2}|\mathbf{p}^{(*)}_i|{\mathrm d}\Omega^{(*)}_i\frac{|\mathbf{p}^{(*)}_{n-1}|{\mathrm d}\Omega^{(*)}_{n-1}}{m_{(n-1)}} \left(\prod_{k=1}^{n-2}\frac{ m_{(k+1)}\mathrm{d}m_{(k+1)}}{ m_{(k)}}\right) \notag \\
     &= \frac1{2^n(2\pi)^{3n}m}\prod_{i=1}^{n-1}|\mathbf{p}^{(*)}_i|{\mathrm d}\Omega^{(*)}_i\left(\prod_{k=2}^{n-1} \mathrm{d}m_{(k)}\right)
     \label{PhaseSpaceHD}
\end{align}
At last, using Eq.~\eqref{PhaseSpaceHD}, we can rewrite the phase-space factor in Eq.~\eqref{DecayFormulaPDG} as follows:
\begin{eqnarray}
\mathrm{d}\Gamma = \frac{(2\pi)^{4-3 n}}{2^{n+1} m^2} |{\cal M}|^2 |{\mathbf p}^{(*)}_1|\cdots|{\mathbf p}^{(*)}_{n-1}|
                   \mathrm{d}\Omega^{(*)}_1\cdots\mathrm{d}\Omega^{(*)}_{n-1}\mathrm{d}m_{(2)}\cdots\mathrm{d}m_{(n-1)}
\label{DecayFormual}
\end{eqnarray}
where $(|{\mathbf p}^{(*)}_k|,\Omega^{(*)}_k)$ is the momentum of particle $k$ in the CM frame of the particle system $k,k+1,\cdots,n$, and
$m_{(k)}$ is the invariant mass of the particle system $k,k+1,\cdots,n$.

\bibliography{JPsiToEtaPiPhi}

\end{document}